\documentclass[10pt,conference,letterpaper,twocolumn]{IEEEtran}

\usepackage[usenames,dvipsnames]{color}

\usepackage{latexsym}
\usepackage{amsxtra} 
\usepackage{amssymb}
\usepackage{amsmath}
\usepackage{pslatex}
\usepackage{epsfig}
\usepackage{wrapfig}
\usepackage{paralist}
\usepackage{stmaryrd}
\usepackage{txfonts}
\usepackage{framed}
\usepackage{makecell}
\usepackage{url}
\usepackage{tikz}
\usetikzlibrary{automata,positioning, calc}
\usepackage[inline,shortlabels]{enumitem}

\usepackage{proof}

\usepackage{algorithm}
\usepackage{algorithmicx}
\usepackage[noend]{algpseudocode}

\usepackage[draft]{commenting}

\newtheorem{lemma}{Lemma}
\newtheorem{theorem}{Theorem}
\newtheorem{proposition}{Proposition}

\newcommand{\set}[1]{\left\{ #1 \right\}}
\newcommand{\tuple}[1]{\left\langle #1 \right\rangle}
\renewcommand{\vec}[1]{\mathbf #1}

\newcommand{\sig}{\mathsf{S}}

\newcommand{\len}[1]{{|{#1}|}}

\newcommand{\nat}{{\bf \mathbb{N}}}
\newcommand{\zed}{{\bf \mathbb{Z}}}

\newcommand{\proj}[2]{{#1}\!\!\downarrow_{{#2}}}

\renewcommand{\paragraph}[1]{\noindent{\bf #1}}

\newif\ifLongVersion\LongVersiontrue
\newcommand{\proof}[1]{\noindent {\em Proof}: {#1}}
\newcommand{\qed}{\hfill$\Box$}

\newcommand{\ssorts}[1]{#1^\mathrm{s}}
\newcommand{\sfuns}[1]{#1^\mathrm{f}}

\newcommand{\iffeq}{\Leftrightarrow}

\newcommand{\teq}{\approx}

\newcommand{\I}{\mathcal{I}}

\newcommand{\fv}[2]{\mathrm{FV}^{#1}(#2)}

\newcommand{\vars}{\mathsf{Var}}
\newcommand{\preds}{\mathsf{\Pi}}

\newcommand{\Data}{\mathsf{Data}}
\newcommand{\Bool}{\mathsf{Bool}}
\newcommand{\booli}{\mathbb{B}}
\newcommand{\true}{\mathtt{true}}
\newcommand{\false}{\mathtt{false}}
\newcommand{\Form}{\mathsf{Form}}

\newcommand{\theory}{\mathbb{T}}

\newcommand{\A}{\mathcal{A}}
\newcommand{\Post}[1]{\mathsf{Post}_{#1}}
\newcommand{\Accept}[1]{\mathsf{Acc}_{#1}}
\newcommand{\AbsPost}[1]{\abs{\mathsf{Post}}_{#1}}
\newcommand{\AbsAccept}[1]{\abs{\mathsf{Acc}}_{#1}}
\newcommand{\Inv}{\mathsf{I}}
\newcommand{\abs}[1]{{#1}^\sharp}
\newcommand{\Art}{\mathcal{T}}
\newcommand{\SubArt}{\mathcal{S}}
\newcommand{\worklist}{\mathtt{WorkList}}
\newcommand{\rootNode}{\mathtt{r}}
\newcommand{\pivot}[1]{\Psi(#1)}
\newcommand{\impact}{\textsc{Impact}}

\begin{document}

\title{The Impact of Alternation}

\author{Radu Iosif and Xiao Xu \\
CNRS, Verimag, Universit\'e de Grenoble Alpes \\
Email: \{Radu.Iosif,Xiao.Xu\}@univ-grenoble-alpes.fr}

\maketitle

\begin{abstract}
Alternating automata have been widely used to model and verify systems
that handle data from finite domains, such as communication protocols
or hardware. The main advantage of the alternating model of
computation is that complementation is possible in linear time, thus
allowing to concisely encode trace inclusion problems that occur often
in verification. In this paper we consider alternating automata over
infinite alphabets, whose transition rules are formulae in a combined
theory of booleans and some infinite data domain, that relate past and
current values of the data variables. The data theory is not fixed,
but rather it is a parameter of the class. We show that union,
intersection and complementation are possible in linear time in this
model and, though the emptiness problem is undecidable, we provide two
efficient semi-algorithms, inspired by two state-of-the-art
abstraction refinement model checking methods: lazy predicate
abstraction \cite{HJMS02} and the \impact~ semi-algorithm
\cite{mcmillan06}. We have implemented both methods and report the
results of an experimental comparison.
\end{abstract}

\section{Introduction}

The language inclusion problem is recognized as being central to
verification of hardware, communication protocols and software
systems. A property is a specification of the correct executions of a
system, given as a set $\mathcal{P}$ of executions, and the
verification problem asks if the set $\mathcal{S}$ of executions of
the system under consideration is contained within $\mathcal{P}$.

This problem is at the core of widespread verification techniques,
such as automata-theoretic model checking \cite{VARDI94}, where
systems are specified as finite-state automata and properties defined
using Linear Temporal Logic \cite{Pnueli77}. However the bottleneck of
this and other related verification techniques is the intractability
of language inclusion (PSPACE-complete for finite-state automata on
finite words over finite alphabets).

Alternation \cite{ChandraKozenStockmeyer81} was introduced as a
generalization of nondeterminism, introducing universal, in addition
to existential transitions. For automata over finite alphabets, the
language inclusion problem can be encoded as the emptiness problem of
an alternating automaton of linear size. Moreover, efficient
exploration techniques based on antichains are shown to perform well
for alternating automata over finite alphabets \cite{DeWulf08}.

Using finite alphabets for the specification of properties and models
is however very restrictive, when dealing with real-life computer
systems, mostly because of the following reasons. On one hand,
programs handle data from very large domains, that can be assumed to
be infinite (64-bit integers, floating point numbers, strings of
characters, etc.) and their correctness must be specified in terms of
the data values. On the other hand, systems must respond to strict
deadlines, which requires temporal specifications as timed
languages \cite{AlurDill94}.

Although being convenient specification tools, automata over infinite
alphabets lack the decidability properties ensured by finite
alphabets. In general, when considering infinite data as part of the
input alphabet, language inclusion is undecidable and, even
complementation becomes impossible, for instance, for timed automata
\cite{AlurDill94} or finite-memory register automata
\cite{Kaminski94}. In some cases, one can recover theoretical
decidability, by restricting the number of variables (clocks) in timed
automata to one \cite{OuaknineWorrell04}, or forbidding relations
between current and past/future values, as with symbolic automata
\cite{symbTransd:POPL12}. In such cases, the emptiness problem for the
alternating versions becomes decidable \cite{Lasota05,DAntoniKW16}.

In this paper, we present a new model of alternating automata over
infinite alphabets consisting of pairs $(a,\nu)$ where $a$ is an input
event from a finite set and $\nu$ is a valuation of a finite set
$\vec{x}$ of variables that range over an infinite domain. We assume
that, at all times, the successive values taken by the variables in
$\vec{x}$ are an observable part of the language, in other words,
there are no hidden variables in our model. The transition rules are
specified by a set of formulae, in a combined first-order theory of
boolean control states and data, that relate past with present values
of the variables. We do not fix the data theory a~priori, but rather
consider it to be a parameter of the class.

A run over an input word $(a_1,\nu_1) \ldots (a_n,\nu_n)$ is a
sequence $\phi_0(\vec{x}_0) \Rightarrow \phi_1(\vec{x}_0,\vec{x}_1)
\Rightarrow \ldots \Rightarrow \phi_n(\vec{x}_0,\ldots,\vec{x}_n)$ of
rewritings of the initial formula by substituting boolean states with
time-stamped transition rules. The word is accepted if the final
formula $\phi_n(\vec{x}_0,\ldots,\vec{x}_n)$ holds, when all
time-stamped variables $\vec{x}_1,\ldots,\vec{x}_n$ are substituted by
their values in $\nu_1,\ldots,\nu_n$, all non-final states replaced by
false and all final states by true.

The boolean operations of union, intersection and complement can be
implemented in linear time in this model, thus matching the complexity
of performing these operations in the finite-alphabet case. The price
to be paid is that emptiness becomes undecidable, for which reason we
provide two efficient semi-algorithms for emptiness, based on lazy
predicate abstraction \cite{HJMS02} and the \impact~method
\cite{mcmillan06}.  These algorithms are proven to terminate and
return a word from the language of the automaton, if one exists, but
termination is not guaranteed when the language is empty.


We have implemented the boolean operations and emptiness checking
semi-algorithms and carried out experiments with examples taken from
array logics \cite{cav09}, timed automata
\cite{henzinger:RealTimeSystems}, communication protocols \cite{abp}
and hardware verification \cite{smrcka}.

\paragraph{Related Work}
Data languages and automata have been defined previously, in a
classical nondeterministic setting. For instance, Kaminski and Francez
\cite{Kaminski94} consider languages, over an infinite alphabet of
data, recognized by automata with a finite number of registers, that
store the input data and compare it using equality. Just as the timed
languages recognized by timed automata \cite{AlurDill94}, these
languages, called quasi-regular, are not closed under complement, but
their emptiness is decidable. The impossibility of complementation
here is caused by the use of hidden variables, which we do not
allow. Emptiness is however undecidable in our case, mainly because
counting (incrementing and comparing to a constant) data values is
allowed, in many data theories.


Another related model is that of predicate automata \cite{Farzan15},
which recognize languages over integer data by labeling the words with
conjunctions of uninterpreted predicates. We intend to explore further
the connection with our model of alternating data automata, in order
to apply our method to the verification of parallel programs.

The model presented in this paper stems from the language inclusion
problem considered in \cite{Tacas16}. There we provide a
semi-algorithm for inclusion of data languages, based on an
exponential determinization procedure and an abstraction refinement
loop using lazy predicate abstraction \cite{HJMS02}. In this work we
consider the full model of alternation and rely entirely on the
ability of SMT solvers to produce interpolants in the combined theory
of booleans and data. Since determinisation is not needed and
complementation is possible in linear time, the bulk of the work is
carried out by the decision procedure.

The emptiness check for alternating data automata adapts similar
semi-algorithms for nondeterministic infinite-state programs to the
alternating model of computation. In particular, we considered the
state-of-the-art \impact~ procedure \cite{mcmillan06} that is shown to
outperform lazy predicate abstraction \cite{HJMS02} in the
nondeterministic case, and generalized it to cope with
alternation. More recent approaches for interpolant-based abstraction
refinement target Horn systems \cite{McMillan14,Hoder12}, used to
encode recursive and concurrent programs
\cite{Grebenshchikov12}. However, the emptiness of alternating word
automata cannot be directly encoded using Horn clauses, because all
the branches of the computation synchronize on the same input, which
cannot be encoded by a finite number of local (equality)
constraints. We believe that the lazy annotation techniques for Horn
clauses are suited for branching computations, which we intend to
consider in a future tree automata setting.

\section{Preliminaries}

A \emph{signature} $\sig = (\ssorts{\sig},\sfuns{\sig})$ consists of a
set $\ssorts{\sig}$ of \emph{sort symbols} and a set $\sfuns{\sig}$ of
sorted \emph{function symbols}. To simplify the presentation, we
assume w.l.o.g. that $\ssorts{\sig} = \set{\Data,\Bool}$\footnote{The
  generalization to more than two sorts is without difficulty, but
  would unnecessarily clutter the technical presentation.} and each
function symbol $f \in \sfuns{\sig}$ has $\#(f) \geq 0$ arguments of
sort $\Data$ and return value $\sigma(f) \in \ssorts{\sig}$. If
$\#(f)=0$ then $f$ is a \emph{constant}. We consider the constants
$\top$ and $\bot$ of sort $\Bool$.

We consider an infinite countable set of \emph{variables} $\vars$,
where each $x \in \vars$ has an associated sort $\sigma(x)$. A
\emph{term} $t$ of sort $\sigma(t)=S$ is a variable $x \in \vars$
where $\sigma(x)=S$, or $f(t_1,\ldots,t_{\#(f)})$ where
$t_1,\ldots,t_{\#(f)}$ are terms of sort $\Data$ and $\sigma(f)=S$. An
\emph{atom} is a term of sort $\Bool$ or an equality $t \teq s$
between two terms of sort $\Data$. A \emph{formula} is an
existentially quantified combination of atoms using disjunction
$\vee$, conjunction $\wedge$ and negation $\neg$ and we write $\phi
\rightarrow \psi$ for $\neg\phi \vee \psi$.

We denote by $\fv{\sigma}{\phi}$ the set of free variables of sort
$\sigma$ in $\phi$ and write $\fv{}{\phi}$ for $\bigcup_{\sigma \in
  \ssorts{\sig}} \fv{\sigma}{\phi}$. For a variable $x \in
\fv{}{\phi}$ and a term $t$ such that $\sigma(t) = \sigma(x)$, let
$\phi[t/x]$ be the result of replacing each occurrence of $x$ by
$t$. For indexed sets $\vec{t}=\set{t_1,\ldots,t_n}$ and
$\vec{x}=\set{x_1,\ldots,x_n}$, we write $\phi[\vec{t}/\vec{x}]$ for
the formula obtained by simultaneously replacing $x_i$ with $t_i$ in
$\phi$, for all $i\in[1,n]$. The size $\len{\phi}$ is the number of
symbols occuring in $\phi$.

An \emph{interpretation} $\I$ maps\begin{inparaenum}[(1)]
\item the sort $\Data$ into a non-empty set $\Data^\I$, 
\item the sort $\Bool$ into the set $\booli = \set{\true,\false}$, where
  $\top^\I = \true$, $\bot^\I = \false$, and
\item each function symbol $f$ into a total function $f^\I :
  (\Data^\I)^{\#(f)} \rightarrow \sigma(f)^I$, or an element of
  $\sigma(f)^I$ when $\#(f)=0$.
\end{inparaenum}
Given an interpretation $\I$, a \emph{valuation} $\nu$ maps each
variable $x \in \vars$ into an element $\nu(x) \in \sigma(x)^\I$. For
a term $t$, we denote by $t^\I_\nu$ the value obtained by replacing
each function symbol $f$ by its interpretation $f^\I$ and each
variable $x$ by its valuation $\nu(x)$. For a formula $\phi$, we write
$\I,\nu \models \phi$ if the formula obtained by replacing each term
$t$ in $\phi$ by the value $t^\I_\nu$ is logically equivalent to true.

A formula $\phi$ is \emph{satisfiable} in the interpretation $\I$ if
there exists a valuation $\nu$ such that $\I,\nu \models \phi$, and
\emph{valid} if $\I,\nu \models \phi$ for all valuations $\nu$.  The
\emph{theory} $\theory(\sig,\I)$ is the set of valid formulae written
in the signature $\sig$, with the interpretation $\I$. A \emph{decision
procedure} for $\theory(\sig,\I)$ is an algorithm that takes a formula
$\phi$ in the signature $\sig$ and returns yes iff $\phi \in
\theory(\sig,\I)$. 

Given formulae $\varphi$ and $\psi$, we say that \emph{$\phi$ entails
  $\psi$}, denoted $\phi \models^\I \psi$ iff $\I,\nu \models \varphi$
implies $\I,\nu \models \psi$, for each valuation $\nu$, and $\phi
\iffeq^\I \psi$ iff $\phi \models^\I \psi$ and $\psi \models^\I \phi$.
We omit mentioning the interpretation $\I$ when it is clear from the
context.

\section{Alternating Data Automata}

In the rest of this section we fix an interpretation $\I$ and a finite
alphabet $\Sigma$ of \emph{input events}. Given a finite set $\vec{x}
\subset \vars$ of variables of sort $\Data$, let $\vec{x} \mapsto
\Data^\I$ be the set of valuations of the variables $\vec{x}$ and
$\Sigma[\vec{x}] = \Sigma \times (\vec{x} \mapsto \Data^\I)$ be the
set of \emph{data symbols}. A \emph{data word} (word in the sequel) is
a finite sequence $(a_1,\nu_1)(a_2,\nu_2) \ldots (a_n,\nu_n)$ of data
symbols, where $a_1,\ldots,a_n \in \Sigma$ and $\nu_1,\ldots,\nu_n :
\vec{x} \rightarrow \Data^\I$ are valuations. We denote by
$\varepsilon$ the empty sequence, by $\Sigma^*$ the set of finite
sequences of input events and by $\Sigma[\vec{x}]^*$ the set of data
words over $\vec{x}$.

This definition generalizes the classical notion of words from a
finite alphabet to the possibly infinite alphabet
$\Sigma[\vec{x}]$. Clearly, when $\Data^\I$ is sufficiently large or
infinite, we can map the elements of $\Sigma$ into designated elements
of $\Data^\I$ and use a special variable to encode the input
events. However, keeping $\Sigma$ explicit in the following simplifies
several technical points below, without cluttering the presentation.

Given sets of variables $\vec{b},\vec{x} \subset \vars$ of sort
$\Bool$ and $\Data$, respectively, we denote by
$\Form(\vec{b},\vec{x})$ the set of formulae $\phi$ such that
$\fv{\Bool}{\phi} \subseteq \vec{b}$ and $\fv{\Data}{\phi} \subseteq
\vec{x}$. By $\Form^+(\vec{b},\vec{x})$ we denote the set of formulae
from $\Form(\vec{b},\vec{x})$ in which each boolean variable occurs
under an even number of negations.

An \emph{alternating data automaton} (ADA or automaton in the sequel)
is a tuple $\A = \tuple{\vec{x},Q,\iota,F,\Delta}$,
where: \begin{compactitem}
\item $\vec{x} \subset \vars$ is a finite set of variables of sort
  $\Data$,
\item $Q \subset \vars$ is a finite set of variables of sort $\Bool$
  (\emph{states}),
\item $\iota \in \Form^+(Q,\emptyset)$ is the \emph{initial
  configuration},
\item $F \subseteq Q$ is a set of \emph{final states}, and
\item $\Delta : Q \times \Sigma \rightarrow
  \Form^+(Q,\overline{\vec{x}}\cup\vec{x})$ is a \emph{transition
    function},
\end{compactitem}
where $\overline{\vec{x}}=\{\overline{x} \mid x \in \vec{x}\}$. In
each formula $\Delta(q,a)$ describing a transition rule, the variables
$\overline{\vec{x}}$ track the previous and $\vec{x}$ the current
values of the variables of $\A$. Observe that the initial values of
the variables are left unconstrained, as the initial configuration
does not contain free data variables. The size of $\A$ is defined as
$\len{\A} = \len{\iota} + \sum_{(q,a) \in
  Q\times\Sigma}\len{\Delta(q,a)}$.

\begin{figure}[htb]
\vspace*{-\baselineskip}
\centerline{\includegraphics[scale=0.8]{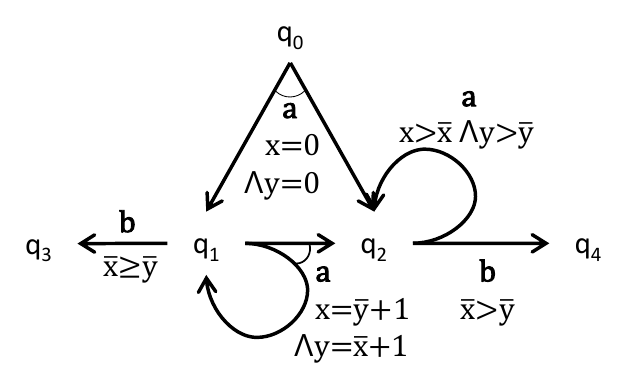}}
\vspace*{-\baselineskip}
\caption{Alternating Data Automaton Example}
\label{fig:ex}
\end{figure}

\paragraph{\em Example}
Figure \ref{fig:ex} depicts an ADA with input alphabet $\Sigma =
\set{a,b}$, variables $\vec{x} = \set{x,y}$, states $Q =
\set{q_0,q_1,q_2,q_3,q_4}$, initial configuration $q_0$, final states
$F=\set{q_3,q_4}$ and transitions:
\[\begin{array}{rcl}
\Delta(q_0,a) & \equiv & q_1 \wedge q_2 \wedge x\teq0 \wedge y\teq0 \\
\Delta(q_1,a) & \equiv & q_1 \wedge q_3 \wedge x\teq\overline{y}+1 \wedge y\teq\overline{x}+1 \\
\Delta(q_1,b) & \equiv & q_3 \wedge \overline{x} \geq \overline{y} \\
\Delta(q_2,a) & \equiv & q_2 \wedge x > \overline{x} \wedge y > \overline{y} \\
\Delta(q_2,b) & \equiv & q_4 \wedge \overline{x}>\overline{y}
\end{array}\]
The missing rules, such as $\Delta(q_0,b)$, are assumed to be
$\bot$. Rules $\Delta(q_0,a)$ and $\Delta(q_1,a)$ are universal and
there are no existential nondeterministic rules. Rules $\Delta(q_1,a)$
and $\Delta(q_2,a)$ compare past ($\overline{x},\overline{y}$) with
present ($x,y$) values, $\Delta(q_0,a)$ constrains the present and
$\Delta(q_1,b)$, $\Delta(q_2,b)$ the past values, respectively. \qed

Formally, let $\vec{x}_k = \set{x_k \mid x \in \vec{x}}$, for any
$k\geq0$, be a set of time-stamped variables. For an input event $a
\in \Sigma$ and a formula $\phi$, we write $\Delta(\phi,a)$
(respectively $\Delta^k(\phi,a)$) for the formula obtained from $\phi$
by simultaneously replacing each state $q \in \fv{\Bool}{\phi}$ by the
formula $\Delta(q,a)$ (respectively
$\Delta(q,a)[\vec{x}_k/\overline{\vec{x}},\vec{x}_{k+1}/\vec{x}]$, for
$k\geq0$). Given a word $w = (a_1,\nu_1)(a_2,\nu_2) \ldots
(a_n,\nu_n)$, the \emph{run} of $\A$ over $w$ is the sequence of
formulae: \[\phi_0(Q) \Rightarrow \phi_1(Q,\vec{x}_0 \cup \vec{x}_1)
\Rightarrow \ldots \Rightarrow \phi_n(Q,\vec{x}_0 \cup \ldots \cup
\vec{x}_n)\] where $\phi_0 \equiv \iota$ and, for all $k\in[1,n]$, we
have $\phi_k \equiv \Delta^k(\phi_{k-1},a_k)$. Next, we slightly abuse
notation and write $\Delta(\iota,a_1,\ldots,a_n)$ for the formula
$\phi_n(\vec{x}_0,\ldots,\vec{x}_n)$ above. We say that $\A$
\emph{accepts} $w$ iff $\I,\nu \models \Delta(\iota,a_1,\ldots,a_n)$,
for some valuation $\nu$ that maps:\begin{inparaenum}[(1)]
\item each $x \in \vec{x}_k$ to $\nu_k(x)$, for all $k\in[1,n]$, 
\item each $q \in \fv{\Bool}{\phi_n} \cap F$ to $\top$ and 
\item each $q \in \fv{\Bool}{\phi_n} \setminus F$ to $\bot$.
\end{inparaenum}
The language of $\A$ is the set $L(\A)$ of words from
$\Sigma[\vec{x}]^*$ accepted by $\A$.

\paragraph{\em Example} 
The following sequence is a non-accepting run of the ADA from Figure
\ref{fig:ex} on the word
$(a,\tuple{0,0}),(a,\tuple{1,1}),(b,\tuple{2,1})$, where $\Data^\I =
\zed$ and the function symbols have standard arithmetic
interpretation:
\[\begin{array}{l}
q_0 \stackrel{(a,\tuple{0,0})}{\Longrightarrow} 
q_1 \wedge q_2 \wedge x_1\teq0 \wedge y_1\teq0 \stackrel{(a,\tuple{1,1})}{\Longrightarrow} \\ 
\underbrace{q_1 \wedge q_2 \wedge x_2\teq y_1+1 \wedge y_2\teq x_1+1}_{q_1} ~\wedge~
\underbrace{q_2 \wedge x_2>x_1 \wedge y_2>y_1}_{q_2} \\ 
\wedge~ x_1\teq0 \wedge y_1\teq0 \stackrel{(b,\tuple{2,1})}{\Longrightarrow} 
\underbrace{q_3 \wedge x_2\geq y_2}_{q_1} \wedge \underbrace{q_4 \wedge x_2>y_2}_{q_2} \wedge x_2\teq y_1+1 \\ 
\wedge~ y_2\teq x_1+1 \wedge \underbrace{q_4 \wedge x_2>y_2}_{q_2} \wedge x_2>x_1 \wedge y_2>y_1 \wedge x_1\teq0 \wedge y_1\teq0 ~\hfill\text{\qed}
\end{array}\]

In this paper we tackle the following problems: \begin{compactenum}
\item \emph{boolean closure}: given automata $\A_1$ and $\A_2$, both
  with the same set of variables $\vec{x}$, do there exist automata
  $\A_\cup$, $\A_\cap$ and $\overline{\A_1}$ such that $L(\A_\cup) =
  \A_1 \cup \A_2$, $L(A_\cap) = \A_1 \cap \A_2$ and
  $L(\overline{\A_1}) = \Sigma[\vec{x}]^* \setminus L(\A_1)$ ?
\item \emph{emptiness}: given an automaton $\A$, is $L(\A) =
  \emptyset$ ?
\end{compactenum}

It is well known that other problems, such as \emph{universality}
(given automaton $\A$ with variables $\vec{x}$, does $L(\A) =
\Sigma[\vec{x}]^*$?) and \emph{inclusion} (given automata $\A_1$ and
$\A_2$ with the same set of variables, does $L(\A_1) \subseteq
L(\A_2)$?) can be reduced to the above problems. Observe furthermore
that we do not consider cases in which the sets of variables in the
two automata differ. An interesting problem in this case would be: given
automata $\A_1$ and $\A_2$, with variables $\vec{x}_1$ and
$\vec{x}_2$, respectively, such that $\vec{x}_1 \subseteq \vec{x}_2$,
does $L(\A_1) \subseteq \proj{L(\A_2)}{\vec{x}_1}$, where
$\proj{L(\A_2)}{\vec{x}_1}$ is the projection of the set of words
$L(\A_2)$ onto the variables $\vec{x}_1$? This problem is considered
as future work.

\subsection{Boolean Closure}

Given a set $Q$ of boolean variables and a set $\vec{x}$ of variables
of sort $\Data$, for a formula $\phi \in \Form^+(Q,\vec{x})$, with no
negated occurrences of the boolean variables, we define the formula
$\overline{\phi} \in \Form^+(Q,\vec{x})$ recursively on the structure
of $\phi$:
\[\begin{array}{lclclcl}
\overline{\phi_1 \vee \phi_2} & \equiv & \overline{\phi_1} \wedge \overline{\phi_2} && 
\overline{\phi_1 \wedge \phi_2} & \equiv & \overline{\phi_1} \vee \overline{\phi_2} \\ 
\overline{\neg\phi} & \equiv & \neg \overline{\phi} \text{ if $\phi$ not atom} &&
\overline{\phi} & \equiv & \phi \text{ if $\phi \in Q$} \\
\overline{\phi} & \equiv & \neg\phi \text{ if $\phi \not\in Q$ atom}
\end{array}\]
We have $\len{\overline{\phi}} = \len{\phi}$, for every formula $\phi
\in \Form^+(Q,\vec{x})$.

In the following let $\A_i =
\tuple{\vec{x},Q_i,\iota_i,F_i,\Delta_i}$, for $i=1,2$, where
w.l.o.g. we assume that $Q_1 \cap Q_2 = \emptyset$. We define:
\[\begin{array}{rcl}
\A_\cup & = & \tuple{\vec{x},Q_1\cup Q_2,\iota_1 \vee \iota_2,F_1\cup F_2,\Delta_1\cup\Delta_2} \\ 
\A_\cap & = & \tuple{\vec{x},Q_1\cup Q_2,\iota_1 \wedge \iota_2, F_1\cup F_2, \Delta_1\cup\Delta_2} \\ 
\overline{\A_1} & = & \langle \vec{x},Q_1,\overline{\iota_1},Q_1\setminus F_1,\overline{\Delta_1} \rangle
\end{array}\] 
where $\overline{\Delta_1}(q,a) \equiv \overline{\Delta_1(q,a)}$, for
all $q \in Q_1$ and $a \in \Sigma$. The following lemma shows the
correctness of the above definitions:

\begin{lemma}\label{lemma:closure}
  Given automata $\A_i = \tuple{\vec{x},Q_i,\iota_i,F_i,\Delta_i}$,
  for $i=1,2$, such that $Q_1 \cap Q_2 = \emptyset$, we have
  $L(\A_\cup) = L(\A_1) \cup L(\A_2)$, $L(\A_\cap) = L(\A_1) \cap
  L(\A_2)$ and $L(\overline{\A_1}) = \Sigma[\vec{x}]^* \setminus
  L(\A_1)$.
\end{lemma}

It is easy to see that $\len{\A_\cup} = \len{\A_\cap} = \len{\A_1} +
\len{\A_2}$ and $\len{\overline{\A}} = \len{\A}$, thus the automata
for the boolean operations, including complementation, can be built in
linear time. This matches the linear-time bounds for intersection and
complementation of alternating automata over finite alphabets
\cite{ChandraKozenStockmeyer81}.

\section{Antichains and Interpolants for Emptiness}

Unlike the boolean closure properties, showed to be effectively
decidable (Lemma \ref{lemma:closure}), the emptiness problem is
undecidable, even in very simple cases. For instance, if $\Data^\I$ is
the set of positive integers, an ADA can simulate an Alternating
Vector Addition System with States (AVASS) using only atoms $x \geq k$
and $x = \overline{x} + k$, for $k \in \zed$, with the classical
interpretation of the function symbols on integers. Since reachability
of a control state is undecidable for AVASS \cite{LINCOLN92}, the
emptiness problem is undecidable for ADA.

Consequently, we give up on the guarantee for termination and build
semi-algorithms that meet the requirements below: \begin{compactenum}[(i)]
\item given an automaton $\A$, if $L(\A) \neq\emptyset$, the procedure
  will terminate and return a word $w \in L(\A)$, and
\item if the procedure terminates without returning such a word, then
  $L(\A) = \emptyset$.
\end{compactenum}

Let us fix an automaton $\A = \tuple{\vec{x},Q,\iota,F,\Delta}$ whose
(finite) input event alphabet is $\Sigma$, for the rest of this
section. Given a formula $\phi \in \Form^+(Q,\vec{x})$ and an input
event $a \in \Sigma$, we define the \emph{post-image} function
$\Post{\A}(\phi,a) \equiv \exists \overline{\vec{x}} ~.~
\Delta(\phi[\overline{\vec{x}}/\vec{x}], a) \in \Form^+(Q,\vec{x})$,
mapping each formula in $\Form^+(Q,\vec{x})$ to a formula defining the
effect of reading the event $a$.

We generalize the post-image function to finite sequences of input
events, as follows:
\[\begin{array}{l}
\Post{\A}(\phi,\varepsilon) \equiv \phi \hspace*{5mm} \Post{\A}(\phi,ua) \equiv \Post{\A}(\Post{\A}(\phi,u),a) \\
\Accept{\A}(u) \equiv \Post{\A}(\iota,u) \wedge \bigwedge_{q \in Q \setminus F} (q \rightarrow \bot)\text{, for any $u \in \Sigma^*$}
\end{array}\]
Then the emptiness problem for $\A$ becomes: does there exist $u \in
\Sigma^*$ such that the formula $\Accept{\A}(u)$ is satisfiable?
Observe that, since we ask a satisfiability query, the final states of
$\A$ need not be constrained\footnote{ Since each state occurs
  positively in $\Accept{\A}(u)$, this formula has a model iff it has
  a model with every $q \in F$ set to true.}. A na\"ive semi-algorithm
enumerates all finite sequences and checks the satisfiability of
$\Accept{\A}(u)$ for each $u \in \Sigma^*$, using a decision procedure
for the theory $\theory(\sig,\I)$.

Since no boolean variable from $Q$ occurs under negation in $\phi$, it
is easy to prove the following monotonicity property: given two
formulae $\phi,\psi \in \Form^+(Q,\vec{x})$ if $\phi \models \psi$
then $\Post{\A}(\phi,u) \models \Post{\A}(\psi,u)$, for any $u \in
\Sigma^*$. This suggest an improvement of the above semi-algorithm,
that enumerates and stores only a set $U \subseteq \Sigma^*$ for which
$\set{\Post{\A}(\phi,u) \mid u \in U}$ forms an
\emph{antichain}\footnote{Given a partial order $(D,\preceq)$ an
  antichain is a set $A \subseteq D$ such that $a \not\preceq b$ for
  any $a,b \in A$.} w.r.t. the entailment partial order. This is
because, for any $u,v \in \Sigma^*$, if $\Post{\A}(\iota,u) \models
\Post{\A}(\iota,v)$ and $\Accept{\A}(uw)$ is satisfiable for some $w
\in \Sigma^*$, then $\Post{\A}(\iota,uw) \models \Post{\A}(\iota,vw)$,
thus $\Accept{\A}(vw)$ is satisfiable as well, and there is no need
for $u$, since the non-emptiness of $\A$ can be proved using $v$
alone. However, even with this optimization, the enumeration of
sequences from $\Sigma^*$ diverges in many real cases, because
infinite antichains exist in many interpretations, e.g.\ $q \wedge x
\teq 0,~ q \wedge x \teq 1, \ldots$ for $\Data^\I = \nat$.

A \emph{safety invariant} for $\A$ is a function $\Inv : (Q \mapsto
\booli) \rightarrow 2^{\vec{x} \mapsto \Data^\I}$ such that, for every
boolean valuation $\beta : Q \rightarrow \booli$, every valuation $\nu
: \vec{x} \mapsto \Data^\I$ of the data variables and every finite
sequence $u \in \Sigma^*$ of input events, the following
hold: \begin{compactenum}
%
\item\label{it1:inv} $\I,\beta \cup \nu \models \Post{\A}(\iota,u)
  \Rightarrow \nu \in \Inv(\beta)$, and
\item\label{it2:inv} $\nu \in \Inv(\beta) \Rightarrow \I,\beta \cup
  \nu \not\models \Accept{\A}(u)$.
\end{compactenum}
If $\Inv$ satisfies only the first point above, we call it an
\emph{invariant}. Intuitively, a safety invariant maps every boolean
valuation into a set of data valuations, that contains the initial
configuration $\iota \equiv \Post{\A}(\iota,\varepsilon)$, whose data
variables are unconstrained, over-approximates the set of reachable
valuations (point \ref{it1:inv}) and excludes the valuations satisfying the
acceptance condition (point \ref{it2:inv}).  A formula $\phi(Q,\vec{x})$
is said to \emph{define} $\Inv$ iff for all $\beta: Q \rightarrow \booli$ and
$\nu : \vec{x} \rightarrow \Data^\I$, we have $\I,\beta\cup\nu \models
\phi$ iff $\nu \in \Inv(\beta)$.

\begin{lemma}\label{lemma:safety-invariant}
  For any automaton $\A$, we have $L(\A) = \emptyset$ if and only if
  $\A$ has a safety invariant.
\end{lemma}

Turning back to the issue of divergence of language emptiness
semi-algorithms in the case $L(\A) = \emptyset$, we can observe that
an enumeration of input sequences $u_1,u_2,\ldots \in \Sigma^*$ can
stop at step $k$ as soon as $\bigvee_{i=1}^k \Post{\A}(\iota,u_i)$
defines a safety invariant for $\A$. Although this condition can be
effectively checked using a decision procedure for the theory
$\theory(\sig,\I)$, there is no guarantee that this check will ever
succeed.

The solution we adopt in the sequel is abstraction to ensure the
termination of invariant computations. However, it is worth pointing
out from the start that abstraction alone will only allow us to build
invariants that are not necessarily safety invariants. To meet the
latter condition, we resort to counterexample guided abstraction
refinement (CEGAR).

Formally, we fix a set of formulae $\preds \subseteq
\Form(Q,\vec{x})$, such that $\bot \in \preds$ and refer to these
formulae as \emph{predicates}. Given a formula $\phi$, we denote by
$\abs{\phi} \equiv \bigwedge \set{\pi \in \preds \mid \phi \models
  \pi}$ the abstraction of $\phi$ w.r.t. the predicates in
$\preds$. The abstract versions of the post-image and acceptance
condition are defined as follows:
\[\begin{array}{l}
\AbsPost{\A}(\phi,\varepsilon) \equiv \phi \hspace*{5mm}
\AbsPost{\A}(\phi,ua) \equiv \abs{(\Post{\A}(\AbsPost{\A}(\phi,u),a))} \\
\AbsAccept{\A}(u) \equiv \AbsPost{\A}(\iota,u) \wedge \bigwedge_{q \in Q \setminus F} 
(q \rightarrow \bot)\text{, for any $u \in \Sigma^*$}
\end{array}\]

\begin{lemma}\label{lemma:abstract-invariant}
  For any bijection $\mu : \nat \rightarrow \Sigma^*$, there exists
  $k>0$ such that $\bigvee_{i=0}^k \AbsPost{\A}(\iota,\mu(i))$ defines
  an invariant $\abs{\Inv}$ for $\A$. 
\end{lemma}


We are left with fulfilling point (\ref{it2:inv}) from the definition
of a safety invariant. To this end, suppose that, for a given set
$\preds$ of predicates, the invariant $\abs{\Inv}$, defined by the
previous lemma, meets point (\ref{it1:inv}) but not point
(\ref{it2:inv}), where $\Post{\A}$ and $\Accept{\A}$ replace
$\AbsPost{\A}$ and $\AbsAccept{\A}$, respectively. In other words,
there exists a finite sequence $u \in \Sigma^*$ such that $\nu \in
\abs{\Inv}(\beta)$ and $\I,\beta\cup\nu \models \AbsAccept{\A}(u)$,
for some boolean $\beta : Q \rightarrow \booli$ and data $\nu :
\vec{x} \rightarrow \Data^\I$ valuations. Such a $u\in\Sigma^*$ is
called a \emph{counterexample}.

Once a counterexample $u$ is discovered, there are two
possibilities. Either\begin{inparaenum}[(i)]
\item $\Accept{\A}(u)$ is satisfiable, in which case $u$ is
  \emph{feasible} and $L(\A) \neq \emptyset$, or
\item $\Accept{\A}(u)$ is unsatisfiable, in which case $u$ is
  \emph{spurious}.
\end{inparaenum}
In the first case, our semi-algorithm stops and returns a witness for
non-emptiness, obtained from the satisfying valuation of
$\Accept{\A}(u)$ and in the second case, we must strenghten the
invariant by excluding from $\abs{\Inv}$ all pairs $(\beta,\nu)$ such
that $\I,\beta\cup\nu \models \AbsAccept{\A}(u)$. This strenghtening
is carried out by adding to $\preds$ several predicates that are
sufficient to exclude the spurious counterexample.

In general, given an unsatisfiable conjunction $\Phi \equiv
\phi_1(X_0,X_1) \wedge \phi_2(X_1,X_2) \wedge \ldots \wedge
\phi_n(X_{n-1},X_n)$ of time-stamped variables $X_i = \set{x_i \mid x
  \in X}$ of any sort, a solution of the \emph{interpolation problem}
$\Phi$, simply called an \emph{interpolant}, is a tuple $\tuple{I_0(X),
  I_1(X), \ldots, I_n(X)}$ such that:\begin{inparaenum}[(i)]
\item $I_0 \equiv \top$,  
\item $I_{i-1}[X_{i-1}/X] \wedge \phi_i(X_{i-1},X_i) \models
  I_i[X_i/X]$, for all $i \in [1,n]$, and
\item $I_n \equiv \bot$. 
\end{inparaenum}
In the following, we shall assume the existence of an interpolating
decision procedure for $\theory(\sig,\I)$.


A classical method for abstraction refinement is to add the elements
of the interpolant obtained from a proof of spuriousness to the set of
predicates. This guarantees progress, meaning that the particular
spurious counterexample, from which the interpolant was generated,
will never be revisited in the future. Though not always, in many
practical test cases this progress property eventually yields a safety
invariant.

Given a non-empty spurious counterexample $u = a_1\ldots a_n$, where
$n>0$, we consider the following interpolation problem: 
\begin{eqnarray}\label{eq:interpolation-problem}
\Theta(u) & \equiv & \theta_0(Q_0) \wedge \theta_1(Q_0 \cup Q_1,\vec{x}_0
\cup \vec{x}_1) \wedge \ldots \\ 
&& \wedge~ \theta_n(Q_{n-1} \cup Q_n,\vec{x}_{n-1} \cup \vec{x}_n) \wedge \theta_{n+1}(Q_n) \nonumber
\end{eqnarray}
where $Q_k = \set{q_k \mid q \in Q}$, $k \in [0,n]$ are time-stamped
sets of boolean variables corresponding to the set $Q$ of states of
$\A$. The first conjunct $\theta_0(Q_0) \equiv \iota[Q_0/Q]$ is the
initial configuration of $\A$, with every $q \in \fv{\Bool}{\iota}$
replaced by $q_0$. The definition of $\theta_k$, for all $k\in[1,n]$,
uses \emph{replacement sets} $R_\ell \subseteq Q_\ell$, $\ell\in
[0,n]$, which are defined inductively below: \begin{compactitem}
\item $R_0 = \fv{\Bool}{\theta_0}$,
\item $\theta_\ell \equiv \bigwedge_{q_{\ell-1}\in R_{\ell-1}}
  (q_{\ell-1} \rightarrow
  \Delta(q,a_\ell)[Q_\ell/Q,\vec{x}_{\ell-1}/\overline{\vec{x}},\vec{x}_\ell/\vec{x}])$
  and $R_\ell = \fv{\Bool}{\theta_\ell} \cap Q_\ell$, for each
  $\ell\in[1,n]$.
\item $\theta_{n+1}(Q_n) \equiv \bigwedge_{q \in Q \setminus F} (q_n
  \rightarrow \bot)$.
\end{compactitem}
The intuition is that $R_0,\ldots,R_n$ are the sets of states
replaced, $\theta_0, \ldots, \theta_n$ are the sets of transition
rules fired on the run of $\A$ over $u$ and $\theta_{n+1}$ is the
acceptance condition, which forces the last remaining non-final states
to be false.

Moreover, we require that an interpolant
$\tuple{\top,I_0,\ldots,I_n,\bot}$ for the interpolation problem
$\Theta(u)$ does not have negative occurrences of states, i.e.\ $I_i
\in \Form^+(Q,\vec{x})$, for all $i \in [0,n]$. Such an interpolant
can always be built, as showed below:

\begin{proposition}\label{prop:positive-interpol}
  If $\Theta(a_1 \ldots a_n)$ is unsatisfiable then one can build an
  interpolant $\tuple{\top,I_0,\ldots,I_n,\bot}$ for $\Theta(a_1\ldots
  a_n)$, such that $I_i \in \Form^+(Q,\vec{x})$, for all $i \in
  [0,n]$.
\end{proposition}

We recall that a run of $\A$ over $u$ is a sequence: \[\phi_0(Q)
\Rightarrow \phi_1(Q,\vec{x}_0\cup\vec{x}_1) \Rightarrow \ldots
\Rightarrow \phi_n(Q,\vec{x}_0\cup\ldots\cup\vec{x}_n)\] where
$\phi_0$ is the initial configuration $\iota$ and for each $k>0$,
$\phi_k$ is obtained from $\phi_{k-1}$ by replacing each state $q \in
\fv{\Bool}{\phi_{k-1}}$ by the formula
$\Delta(q,a_k)[\vec{x}_{k-1}/\overline{\vec{x}},\vec{x}_k/\vec{x}]$,
given by the transition function of $\A$. Observe that, because the
states are replaced with transition formulae when moving one step in a
run, these formulae lose track of the control history and are not
suitable for producing interpolants that relate states and data.

The main idea behind the above definition of the interpolation problem
is that we would like to obtain an interpolant $\tuple{\top,I_0(Q),
  I_1(Q,\vec{x}), \ldots, I_{n}(Q,\vec{x}),\bot}$ whose formulae
\emph{combine states with the data constraints that must hold
  locally}, whenever the control reaches a certain boolean
configuration. This association of states with data valuations is
tantamount to defining efficient semi-algorithms, based on lazy
abstraction \cite{HJMS02}. Furthermore, the abstraction defined by the
interpolants generated in this way can also \emph{over-approximate the
  control structure} of an automaton, in addition to the sets of data
values encountered throughout its runs.

The correctness of this interpolation-based abstraction refinement
setup is captured by the progress property below, which guarantees
that adding the formulae of an interpolant for $\Theta(u)$ to the set
$\preds$ of predicates suffices to exclude the spurious counterexample
$u$ from future searches.

\begin{lemma}\label{lemma:progress}
  For any sequence $u = a_1\ldots a_n \in \Sigma^*$, if
  $\Accept{\A}(u)$ is unsatisfiable, the following
  hold: \begin{compactenum}
  \item\label{it1:progress} $\Theta(u)$ is unsatisfiable, and
  \item\label{it2:progress} if $\tuple{\top,I_0,\dots,I_{n},\bot}$ is
    an interpolant for $\Theta(u)$ such that $\set{I_i \mid i
      \in[0,n]} \subseteq \preds$ then $\AbsAccept{\A}(u)$ is
    unsatisfiable.
  \end{compactenum}
\end{lemma}

\section{Lazy Predicate Abstraction for ADA Emptiness}

We have now all the ingredients to describe the first emptiness
checking semi-algorithm for alternating data
automata. Algorithm\footnote{Though termination is not guaranteed, we
  call it algorithm for conciseness.} \ref{alg:predabs} builds an
\emph{abstract reachability tree} (ART) whose nodes are labeled with
formulae over-approximating the concrete sets of configurations, and a
covering relation between nodes in order to ensure that the set of
formulae labeling the nodes in the ART forms an antichain. Any
spurious counterexample is eliminated by computing an interpolant and
adding its formulae to the set of predicates (cf. Lemma \ref{lemma:progress}).

\begin{algorithm}[t!]
{\scriptsize\begin{algorithmic}[0]
\State {\bf input}: an ADA $\A = \tuple{\vec{x},Q,\iota,F,\Delta}$
over the alphabet $\Sigma$ of input events

\State {\bf output}: $\true$ if $L(\A)=\emptyset$ and a data word $w
\in L(\A)$ otherwise
\end{algorithmic}}

{\scriptsize\begin{algorithmic}[1] 

  \State let $\Art = \tuple{N,E,\rootNode,\Lambda,\lhd}$ be an ART

  \State initially $N = E = \lhd = \emptyset$, $\Lambda =
  \set{(\rootNode,\iota)}$, $\preds = \set{\bot}$, $\worklist =
  \tuple{\rootNode}$,

  \While {$\worklist \neq \emptyset$}
  \label{ln:while}
  
  \State dequeue $n$ from $\worklist$ 

  \State $N \leftarrow N \cup \set{n}$

  \State let $\lambda(n) = a_1\ldots a_k$ be the label of the path from
  $\rootNode$ to $n$ 

  \If{$\AbsPost{\A}(\lambda(n))$ is satisfiable} \Comment{counterexample candidate}

  \If{$\Accept{\A}(u)$ is satisfiable} \Comment{feasible counterexample}
  \label{ln:feasible}

  \State get model $(\beta,\nu_1,\ldots,\nu_k)$ of
  $\Accept{\A}(\lambda(n))$

  \State {\bf return} $w = (a_1,\nu_1) \ldots (a_k,\nu_k)$
  \label{ln:cex}
  \Comment{$w \in L(\A)$ by construction}

  \Else \Comment{spurious counterexample}

  \State $p \leftarrow \Call{FindPivot}{\lambda(n),\Art}$
  \label{ln:pivot}

  \State $v \leftarrow \Call{LeastInfeasibleSuffix}{\lambda(n),\Art}$
  
  \State $\preds \leftarrow \preds \cup \set{I_0,\ldots,I_\ell}$,
  where $\tuple{\top,I_0,\ldots,I_\ell,\bot}$ is an interpolant for
  $\pivot{v}$

  \State let $\SubArt = \tuple{N',E',p,\Lambda',\lhd'}$ be the subtree
  of $\Art$ rooted at $p$

  \For{$(m,q) \in \lhd$ such that $q \in N'$}

  \State remove $m$ from $N$ and enqueue $m$ into $\worklist$

  \State remove $\SubArt$ from $\Art$

  \State enqueue $p$ into $\worklist$
  \label{ln:refine}
  \Comment{recompute the subtree rooted at $p$}

  \EndFor

  \EndIf

  \Else \label{ln:else}

  \For{$a \in \Sigma$}\label{ln:forall-events}
  \Comment{expand $n$}

  \State $\phi \leftarrow \AbsPost{\A}(\Lambda(n),a)$

  \If{exist $m \in N$ such that $\phi \models \Lambda(m)$} \label{ln:covered}

  \State $\lhd \leftarrow \lhd \cup \set{(n,m)}$ \Comment{$m$ covers $n$}
  \label{ln:direct-cover}

  \Else 

  \State let $s$ be a fresh node

  \State $E \leftarrow E \cup \set{(n,a,s)}$

  \State $\Lambda \leftarrow \Lambda \cup \set{(s,\phi)}$

  \State $R \leftarrow \set{m \in \worklist \mid \Lambda(m) \models
    \phi}$ \Comment{worklist nodes covered by $s$}

  \For{$r \in R$}
  
  \For{$m \in N$ such that $(m,b,r) \in E$, $b \in \Sigma$}

  \State $\lhd \leftarrow \lhd \cup \set{(m,s)}$ 
  \label{ln:child-cover}
  \Comment{redirect covered children from $R$ into $s$}

  \EndFor

  \For{$(m,r) \in \lhd$}

  \State $\lhd \leftarrow \lhd \cup \set{(m,s)}$ 
  \label{ln:indirect-cover}
  \Comment{redirect covered nodes from $R$ into $s$}

  \EndFor

  \EndFor

  \State remove $R$ from $\Art$

  \State enqueue $s$ into $\worklist$ 
  \label{ln:expand}

  \EndIf

  \EndFor

  \EndIf

  \EndWhile

  \State {\bf return} $\true$
  \label{ln:true}
\end{algorithmic}}
\caption{Lazy Predicate Abstraction for ADA Emptiness}
\label{alg:predabs}
\end{algorithm}

Formally, an ART is tuple $\Art =
\tuple{N,E,\rootNode,\Lambda,R,T,\lhd}$, where: \begin{compactitem}
\item $N$ is a set of nodes, 
\item $E \subseteq N \times \Sigma \times
N$ is a set of edges, 
\item $\rootNode \in N$ is the root of the directed
tree $(N,E)$, 
\item $\Lambda : N \rightarrow \Form(Q,\vec{x})$ is a labeling of the
  nodes with formulae, such that $\Lambda(\rootNode) = \iota$,
\item $R : N \rightarrow 2^Q$ is a labeling of nodes with replacement
  sets, such that $R(\rootNode) = \fv{\Bool}{\iota}$,
\item $T : E \rightarrow \bigcup_{i=0}^\infty
  \Form^+(Q_i,\vec{x}_i,Q_{i+1},\vec{x}_{i+1})$ is a labeling of edges
  with time-stamped formulae, and
\item $\lhd \subseteq N \times N$ is a set of
\emph{covering edges}. 
\end{compactitem}

Each node $n \in N$ corresponds to a unique path from the root to $n$,
labeled by a sequence $\lambda(n) \in \Sigma^*$ of input events. The
\emph{least infeasible suffix} of $\lambda(n)$ is the smallest
sequence $v = a_1 \ldots a_k$, such that $\lambda(n) = wv$, for some
$w \in \Sigma^*$ and the following formula is unsatisfiable:
\begin{equation}\label{eq:pivot}
  \pivot{v} \equiv \Lambda(p)[Q_0/Q] \wedge
  \theta_1(Q_0 \cup Q_1,\vec{x}_0 \cup \vec{x}_1) \wedge \ldots \wedge
  \theta_{k+1}(Q_k)
\end{equation}
where $\theta_1,\ldots,\theta_{k+1}$ are defined as in
(\ref{eq:interpolation-problem}) and $\theta_0 \equiv
\Lambda(p)[Q_0/Q]$. The \emph{pivot} of $n$ is the node $p$
corresponding to the start of the least infeasible suffix.  We assume
the existence of two functions $\Call{FindPivot}{u,\Art}$ and
$\Call{LeastInfeasibleSuffix}{u,\Art}$ that return the pivot and least
infeasbile suffix of a sequence $u \in \Sigma^*$ in an ART $\Art$,
without detailing their implementation. 

With these considerations, Algorithm \ref{alg:predabs} uses a worklist
iteration to build an ART. We keep newly expanded nodes of $\Art$ in a
queue $\worklist$, thus implementing a breadth-first exploration
strategy, which guarantees that the shortest counterexamples are
explored first. When the search encounters a counterexample candidate
$u$, it is checked for spuriousness. If the counterexample is
feasible, the procedure returns a data word $w \in L(\A)$, which
interleaves the input events of $u$ with the data valuations from the
model of $\Accept{\A}(u)$ (since $u$ is feasible, clearly
$\Accept{\A}(u)$ is satisfiable). Otherwise, if $u$ is spurious, we
compute its pivot $p$ (line \ref{ln:pivot}), add the interpolants for
the least unfeasible suffix of $u$ to the set of predicates $\preds$,
remove and recompute the subtree of $\Art$ rooted at $p$.

Termination of Algorithm \ref{alg:predabs} depends on the ability of a
given interpolating decision procedure for the combined boolean and
data theory $\theory(\sig,\I)$ to provide interpolants that yield a
safety invariant, whenever $L(\A) = \emptyset$. In this case, we use
the covering relation $\lhd$ to ensure that, when a newly generated
node is covered by a node already in $N$, it is not added to the
worklist, thus cutting the current branch of the search. 

Formally, for any two nodes $n,m \in N$, we have $n \lhd m$ iff
$\AbsPost{\A}(\Lambda(n),a) \models \Lambda(m)$ for some $a \in
\Sigma$, in other words, if $n$ has a successor whose label entails
the label of $m$.

\begin{figure}[t!]
\begin{tabular}{cc}
\includegraphics[scale=0.5]{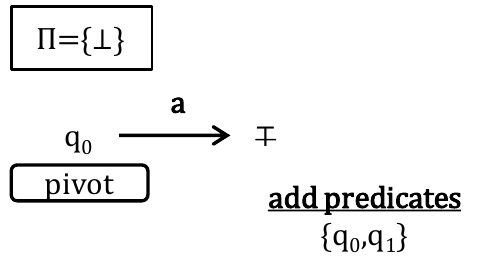} & \includegraphics[scale=0.5]{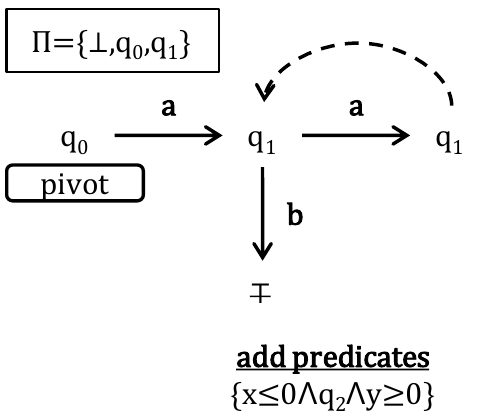} \\[-2mm]
\tiny{(a)} & \tiny{(b)} \\
\\
\includegraphics[scale=0.5]{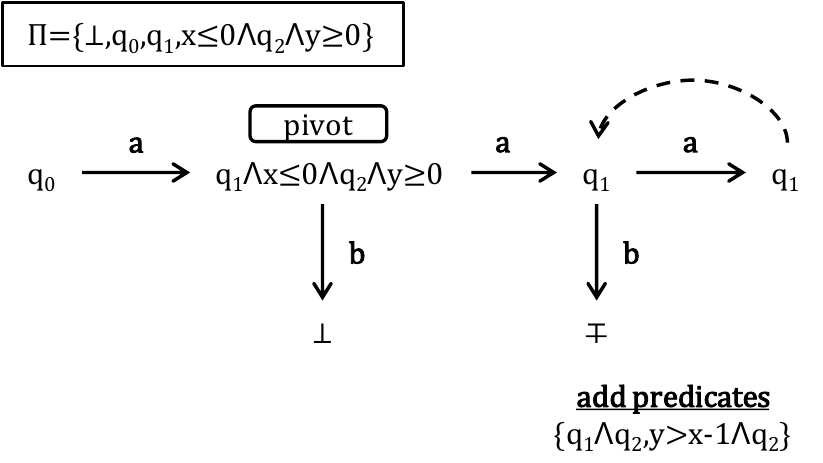} & \includegraphics[scale=0.5]{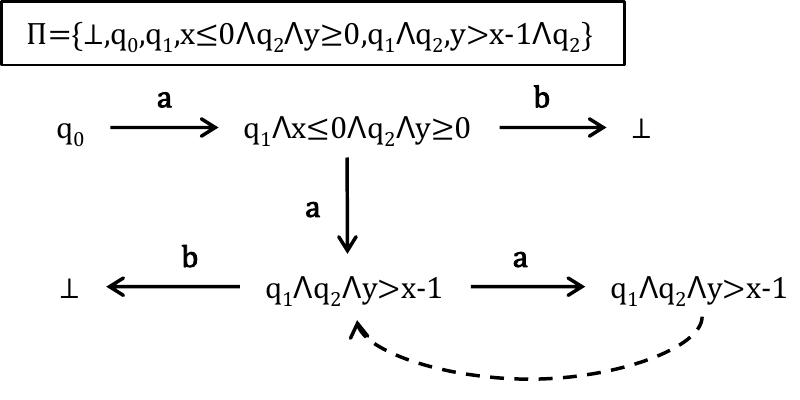} \\[-2mm]
\tiny{(c)} & \tiny{(d)}
\end{tabular}
\vspace*{-\baselineskip}
\caption{Proving Emptiness of the Automaton from Fig. \ref{fig:ex} by Algorithm \ref{alg:predabs}}
\label{fig:pa}
\end{figure}

\paragraph{\em Example}
Consider the automaton given in Figure \ref{fig:ex}. First, Algorithm
\ref{alg:predabs} fires the sequence $a$, and since there are no other
formulae than $\bot$ in $\preds$, the successor of $\iota \equiv q_0$
is $\top$, in Figure \ref{fig:pa} (a). The spuriousness check for $a$
yields the root of the ART as pivot and the interpolant
$\tuple{q_0,q_1}$, which is added to the set $\preds$. Then the $\top$
node is removed and the next time $a$ is fired, it creates a node
labeled $q_1$. The second sequence $aa$ creates a successor node
$q_1$, which is covered by the first, depicted with a dashed arrow, in
Figure \ref{fig:pa} (b). The third sequence is $ab$, which results in
a new uncovered node $\top$ and triggers a spuriousness check. The new
predicate obtained from this check is $x\leq0\wedge q_2 \wedge y\geq0$
and the pivot is again the root. Then the entire ART is rebuilt with
the new predicates and the fourth sequence $aab$ yields an uncovered
node $\top$, in Figure \ref{fig:pa} (c). The new pivot is the endpoint
of $a$ and the newly added predicates are $q_1\wedge q_2$ and $y>x-1
\wedge q_2$. Finally, the ART is rebuilt from the pivot node and
finally all nodes are covered, thus proving the emptiness of the
automaton, in Figure \ref{fig:pa} (d). \qed

The correctness of Algorithm \ref{alg:predabs} is proved below:

\begin{theorem}\label{thm:predabs}
  Given an automaton $\A$, such that $L(\A) \neq \emptyset$, Algorithm
  \ref{alg:predabs} terminates and returns a word $w \in L(\A)$. If
  Algorithm \ref{alg:predabs} terminates reporting $\true$, then
  $L(\A) = \emptyset$.
\end{theorem}

\section{Checking ADA Emptiness with \impact}

As pointed out by a number of authors, the bottleneck of predicate
abstraction is the high cost of reconstructing parts of the ART,
subsequent to the refinement of the set of predicates. The main idea
of the \impact~ procedure \cite{mcmillan06} is that this can be
avoided and the refinement (strenghtening of the node labels of the
ART) can be performed in-place. This refinement step requires an
update of the covering relation, because a node that used to cover
another node might not cover it after the strenghtening of its label.

We consider a total alphabetical order $\prec$ on $\Sigma$ and lift it
to the total lexicographical order $\prec^*$ on $\Sigma^*$. A node $n
\in N$ is \emph{covered} if $(n,p) \in \lhd$ or it has an ancestor $m$
such that $(m,p) \in \lhd$, for some $p \in N$. A node $n$ is
\emph{closed} if it is covered, or $\Lambda(n) \not\models \Lambda(m)$
for all $m \in N$ such that $\lambda(m) \prec^* \lambda(n)$. Observe
that we use the coverage relation $\lhd$ here with a different meaning
than in Algorithm \ref{alg:predabs}.

\begin{algorithm}[t!]
{\scriptsize\begin{algorithmic}[0]
  \State {\bf input}: an ADA $\A = \tuple{\vec{x},Q,\iota,F,\Delta}$
  over the alphabet $\Sigma$ of input events

  \State {\bf output}: $\true$ if $L(\A)=\emptyset$ and a data word $w
  \in L(\A)$ otherwise
\end{algorithmic}}

{\scriptsize\begin{algorithmic}[1] 

  \State let $\Art = \tuple{N,E,\rootNode,\Lambda,R,T,\lhd}$ be an ART

  \State initially $N = E = T = \lhd = \emptyset$, $\Lambda =
  \set{(\rootNode,\iota)}$, $R = \fv{\Bool}{\iota[Q_0/Q]}$,
  $\worklist=\set{\rootNode}$
  
  \While{$\worklist \neq \emptyset$}
  \label{ln:impact-while}

  \State dequeue $n$ from $\worklist$ 
  \label{ln:impact-dequeue}

  \State $N \leftarrow N \cup \set{n}$

  \State let $(\rootNode,a_1,n_1),(n_1,a_2,n_2), \ldots,
  (n_{k-1},a_k,n)$ be the path from $\rootNode$ to $n$

  \If{$\Accept{\A}(a_1\ldots a_k)$ is satisfiable} 
  \Comment{counterexample is feasible}

  \State get model $(\beta,\nu_1,\ldots,\nu_k)$ of
  $\Accept{\A}(\lambda(n))$

  \State {\bf return} $w = (a_1,\nu_1) \ldots (a_k,\nu_k)$
  \label{ln:impact-nonempty}
  \Comment{$w \in L(\A)$ by construction}

  \Else \Comment{spurious counterexample}

  \State let $\tuple{\top,I_0,\ldots,I_k,\bot}$ be an interpolant for
  $\Theta(a_1 \ldots a_k)$
  \label{ln:refine-begin}

  \State $b \leftarrow \false$

  \For{$i=0,\ldots,k$}

  \If{$\Lambda(n_i) \not\models I_i$}

  \State $\lhd \leftarrow \lhd \setminus \set{(m,n_i) \in \lhd \mid m \in N}$

  \State $\Lambda(n_i) \leftarrow \Lambda(n_i) \wedge I_i$
  \Comment{strenghten the label of $n_i$}

  \If{$\neg b$}

  \State $b \leftarrow \Call{Close}{n_i}$
  \label{ln:refine-end}

  \EndIf 

  \EndIf 

  \EndFor 

  \EndIf 

  \If{$n$ is not covered}

  \For{$a \in \Sigma$} 
  \Comment{expand $n$}
  \label{ln:expand-begin}

  \State let $s$ be a fresh node and $e = (n,a,s)$ be a new edge

  \State $E \leftarrow E \cup \set{e}$
  \label{ln:edge-insert}

  \State $\Lambda \leftarrow \Lambda \cup \set{(s,\top)}$

  \State $T \leftarrow T \cup \set{(e, \theta_k)}$

  \State $R \leftarrow R \cup \{(s,\bigcup_{q \in R(n)}\fv{\Bool}{\Delta(q,a)})\}$

  \State enqueue $s$ into $\worklist$
  \label{ln:expand-end}

  \EndFor

  \EndIf 

  \EndWhile  

  \State {\bf return} $\true$
\end{algorithmic}

\begin{algorithmic}[1]
\Function{Close}{$x$} {\bf returns} $\Bool$

\For{$y \in N$ such that $\lambda(y) \prec^* \lambda(x)$}

\If{$\Lambda(x) \models \Lambda(y)$}

\State $\lhd \leftarrow \left(\lhd \setminus \set{ (p,q) \in \lhd \mid
  \text{$q$ is $x$ or a successor of $x$}}\right) \cup \set{(x,y)}$
\label{ln:close-uncover}

\State {\bf return} $\true$ 

\EndIf

\EndFor

\State {\bf return} $\false$

\EndFunction
\end{algorithmic}}
\caption{\impact~ for ADA Emptiness}
\label{alg:impact}
\end{algorithm}

The execution of Algorithm \ref{alg:impact} consists of three
phases\footnote{Corresponding to the \textsc{Close}, \textsc{Refine}
  and \textsc{Expand} in \cite{mcmillan06}.}: \emph{close},
\emph{refine} and \emph{expand}. Let $n$ be a node removed from the
worklist at line \ref{ln:impact-dequeue}. If $\Accept{\A}(\lambda(n))$
is satisfiable, the counterexample $\lambda(n)$ is feasible, in which
case a model of $\Accept{\A}(\lambda(n))$ is obtained and a word $w
\in L(\A)$ is returned. Otherwise, $\lambda(n)$ is a spurious
counterexample and the procedure enters the refinement phase (lines
\ref{ln:refine-begin}-\ref{ln:refine-end}). The interpolant for
$\Theta(\lambda(n))$ (cf. formula \ref{eq:interpolation-problem}) is
used to strenghten the labels of all the ancestors of $n$, by
conjoining the formulae of the interpolant to the existing labels.

In this process, the nodes on the path between $\rootNode$ and $n$,
including $n$, might become eligible for coverage, therefore we
attempt to close each ancestor of $n$ that is impacted by the
refinement (line \ref{ln:refine-end}). Observe that, in this case the
call to $\Call{Close}{}$ must uncover each node which is covered by a
successor of $n$ (line \ref{ln:close-uncover} of the $\Call{Close}{}$
function). This is required because, due to the over-approximation of
the sets of reachable configurations, the covering relation is not
transitive, as explained in \cite{mcmillan06}. If $\Call{Close}{}$
adds a covering edge $(n_i,m)$ to $\lhd$, it does not have to be
called for the successors of $n_i$ on this path, which is handled via
the boolean flag $b$.

Finally, if $n$ is still uncovered (it has not been previously covered
during the refinement phase) we expand $n$ (lines
\ref{ln:expand-begin}-\ref{ln:expand-end}) by creating a new node for
each successor $s$ via the input event $a \in \Sigma$ and inserting it
into the worklist.

\begin{figure}[htb]
\vspace*{-\baselineskip}
\begin{tabular}{c}
\includegraphics[scale=0.65]{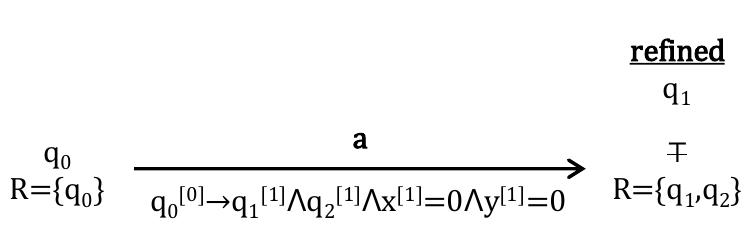} \\
\tiny{(a)} \\
\includegraphics[scale=0.65]{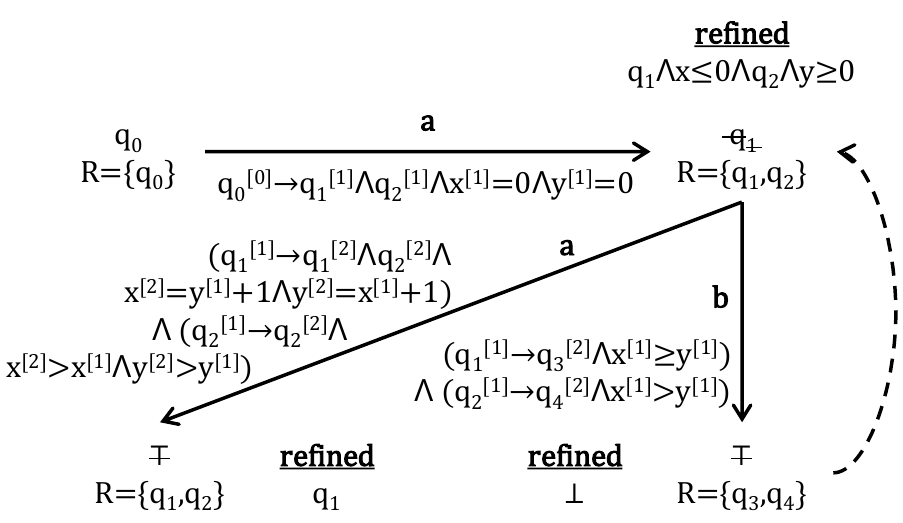} \\[-2mm]
\tiny{(b)} \\
\includegraphics[scale=0.65]{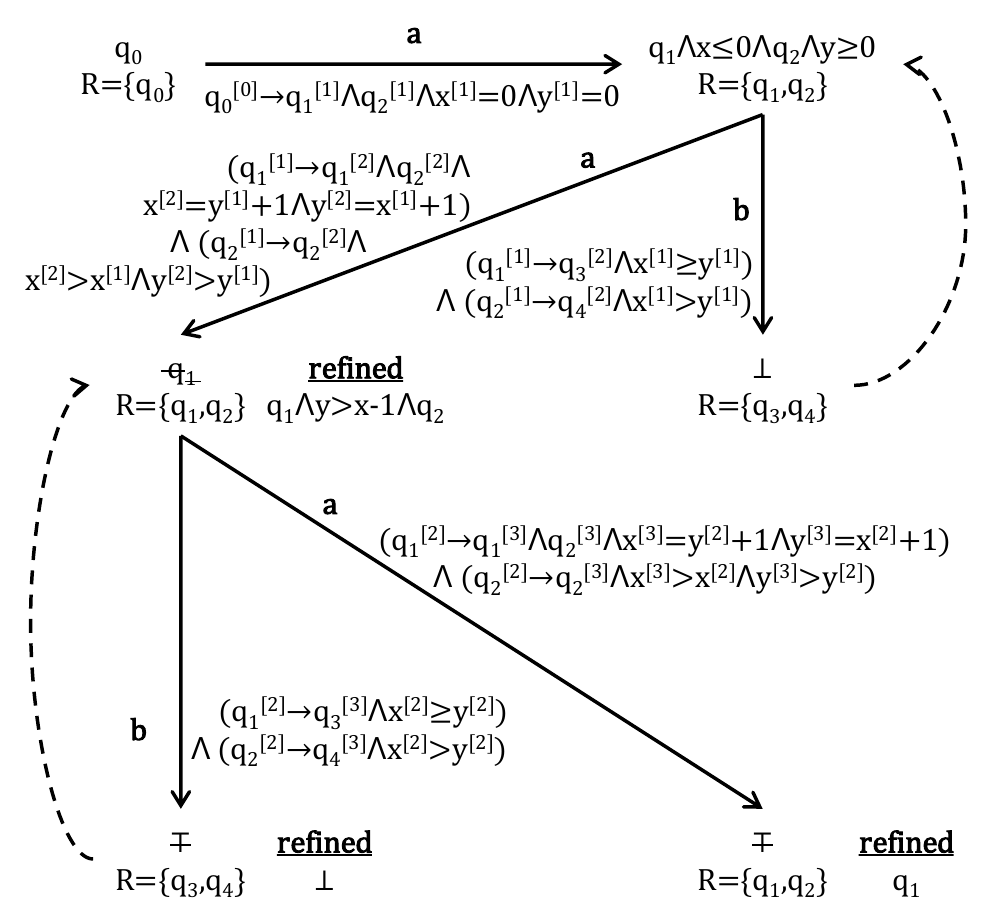} \\[-2mm]
\tiny{(c)}
\end{tabular}
\vspace*{-0.5\baselineskip}
\caption{Proving Emptiness of the Automaton from Fig. \ref{fig:ex}
  by Algorithm \ref{alg:impact} (1/2)}
\label{fig:im1}
\vspace*{-\baselineskip}
\end{figure}

\begin{figure}[htb]
\centerline{\includegraphics[scale=0.65]{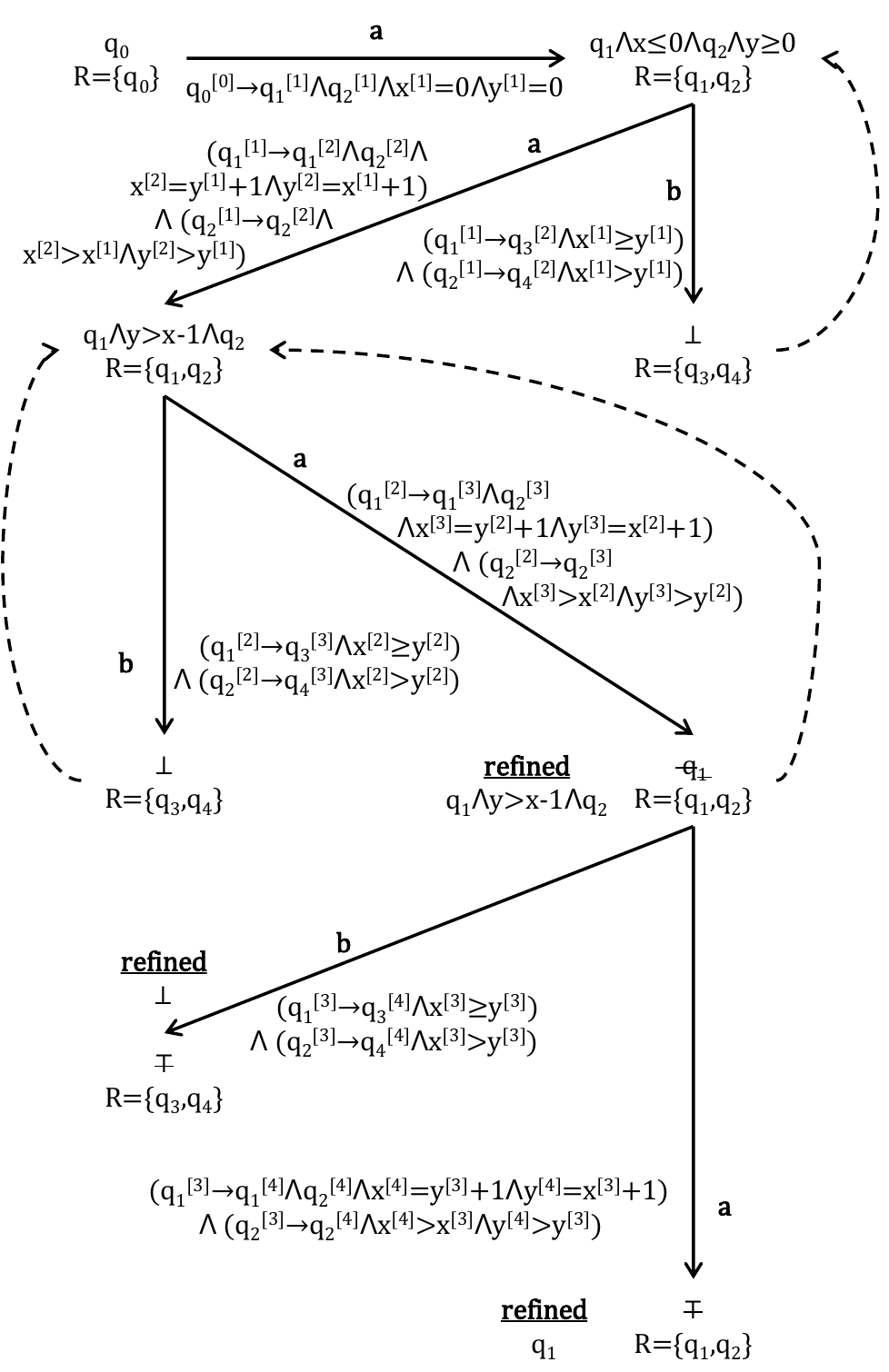}}
\vspace*{-\baselineskip}
\caption{Proving Emptiness of the Automaton from Fig. \ref{fig:ex}
  by Algorithm \ref{alg:impact} (2/2)}
\label{fig:im2}
\end{figure}

\paragraph{\em Example}
We show the execution of Algorithm \ref{alg:impact} on the automaton
from Figure \ref{fig:ex}.  Initially, the procedure fires the sequence
$a$, whose endpoint is labeled with $\top$, in Figure \ref{fig:im1}
(a). Since this node is uncovered, we check the spuriousness of the
counterexample $a$ and refine the label of the node to $q_1$. Since
the node is still uncovered, two successors, labeled with $\top$ are
computed, corresponding to the sequences $aa$ and $ab$, in Figure
\ref{fig:im1} (b). The spuriousness check for $aa$ yields the
interpolant $\tuple{q_0,x\leq0\wedge q_2 \wedge y\geq0}$ which
strenghtens the label of the endpoint of $a$ from $q_1$ to $q_1\wedge
x\leq0\wedge q_2\wedge y\geq0$. The sequence $ab$ is also found to be
spurious, which changes the label of its endpoint from $\top$ to
$\bot$, and also covers it (depicted with a dashed edge). Since the
endpoint of $aa$ is not covered, it is expanded to $aaa$ and $aab$, in
Figure \ref{fig:im1} (c). Both sequences $aaa$ and $aab$ are found to
be spurious, and the enpoint of $aab$, whose label has changed from
$\top$ to $\bot$, is now covered. In the process, the label of $aa$
has also changed from $q_1$ to $q_1 \wedge y>x-1 \wedge q_2$, due to
the strenghtening with the interpolant from $aab$.  Finally, the only
uncovered node $aaa$ is expanded to $aaaa$ and $aaab$, both found to be
spurious, in Figure \ref{fig:im2}. The refinement of $aaab$ causes
the label of $aaa$ to change from $q_1$ to $q_1 \wedge y>x-1 \wedge
q_2$ and this node is now covered by $aa$. Since its successors are
also covered, there are no uncovered nodes and the procedure returns
$\true$. \qed

\begin{theorem}\label{thm:impact}
  Given an automaton $\A$, such that $L(\A) \neq \emptyset$, Algorithm
  \ref{alg:impact} terminates and returns a word $w \in L(\A)$. If
  Algorithm \ref{alg:impact} terminates reporting $\true$, then
  $L(\A) = \emptyset$.
\end{theorem}

\section{Experimental Evaluation}

We have implemented both Algorithm \ref{alg:predabs} and
\ref{alg:impact} in a prototype tool\footnote{The implementation is
  available at \url{https://github.com/cathiec/AltImpact}} that uses
the Z3 SMT solver\footnote{\url{https://github.com/Z3Prover/z3}} for
the satisfiability queries and interpolant generation, in the theory
of linear integer arithmetic (LIA) combined with booleans. We compared
both algorithms with a previous implementation of a trace inclusion
procedure, called
\textsc{Includer}\footnote{\url{http://www.fit.vutbr.cz/research/groups/verifit/tools/includer/}},
that uses on-the-fly determinisation and lazy predicate abstraction
with interpolant-based refinement \cite{Tacas16} in the LIA theory,
without booleans.

The results of the experiments are given in Table
\ref{tab:experiments}. We applied the tool first to several array
logic entailments, which occur as verification conditions for
imperative programs with arrays \cite{cav09} (array\_shift,
array\_simple, array\_rotation 1+2) available online
\cite{ntslib}. Next, we applied it on proving safety properties of
hardware circuits (hw1+2) \cite{smrcka}. Finally, we considered two
timed communication protocols, consisting of systems that are
asynchronous compositions of timed automata, whom correctness
specifications are given by timed automata monitors: a timed version of
the Alternating Bit Protocol (abp) \cite{abp} and a~controller of a
railroad crossing (train) \cite{henzinger:RealTimeSystems}. All
results were obtained on an Intel(R) Core(TM) i7-4650U CPU @ 1.70GHz
with 8GB of RAM. The automata sizes are in bytes and the execution
times are in seconds.

\begin{table}[htb]
\begin{center}
{\fontsize{8}{9}\selectfont
\begin{tabular}{||l|c|c|c|c|c||}
\hline
Example & $\len{\A}$ & $L(\A)=\emptyset$ ? & Algorithm \ref{alg:predabs} & Algorithm \ref{alg:impact} & \textsc{Includer} \\
\hline
simple1 & 309 & no & 0.641 & 0.061 & 0.021\\
\hline
simple2 & 504 & yes & 0.667 & 0.179 & 0.031\\
\hline
simple3 & 214 & yes & 0.825 & 0.150 & 0.051\\
\hline
array\_shift & 874 & yes & 2.138 & 0.209 & 0.060\\
\hline
array\_simple & 3440 & yes & timeout & 16.435 & 5.496\\
\hline
array\_rotation1 & 1834 & yes & 6.827 & 0.865 & 0.114\\
\hline
array\_rotation2 & 15182 & yes & timeout & timeout & 23.056\\
\hline
abp & 6909 & no & 8.292 & 1.410 & 1.715\\
\hline
train & 1823 & yes & 16.989 & 2.738 & 0.308\\
\hline
hw1 & 322 & yes & 1.566 & 0.298 & 0.159\\
\hline
hw2 & 674 & yes & 22.914 & 0.552 & 0.419\\
\hline
\end{tabular}
}
\caption{}\label{tab:experiments}
\end{center}
\end{table}

As in the non-deterministic case \cite{mcmillan06}, \impact~
outperforms lazy predicate abstraction for checking emptiness by at
least one order of magnitude. However, both our implementations are
slower than \textsc{Includer}, on average (except for the abp
example). The reason for this is currently under investigation, one
possible bottleneck being the hardness of the combined (LIA+booleans)
interpolation problems, as opposed to converting the entire formula
into DNF, eliminating the boolean variables and using interpolation in
the pure LIA theory.

\bibliographystyle{abbrv} \bibliography{refs}

\begin{thebibliography}{10}

\bibitem{AlurDill94}
R.~Alur and D.~L. Dill.
\newblock A theory of timed automata.
\newblock {\em Theor. Comput. Sci.}, 126(2):183--235, 1994.

\bibitem{cav09}
M.~Bozga, P.~Habermehl, R.~Iosif, F.~Konecn{\'{y}}, and T.~Vojnar.
\newblock Automatic verification of integer array programs.
\newblock In {\em Proc. of CAV'09}, volume 5643 of {\em LNCS}, pages 157--172,
  2009.

\bibitem{ChandraKozenStockmeyer81}
A.~K. Chandra, D.~C. Kozen, and L.~J. Stockmeyer.
\newblock Alternation.
\newblock {\em J. ACM}, 28(1):114--133, 1981.

\bibitem{DAntoniKW16}
L.~D'Antoni, Z.~Kincaid, and F.~Wang.
\newblock A symbolic decision procedure for symbolic alternating finite
  automata.
\newblock {\em CoRR}, abs/1610.01722, 2016.

\bibitem{DeWulf08}
M.~De~Wulf, L.~Doyen, N.~Maquet, and J.~F. Raskin.
\newblock Antichains: Alternative algorithms for ltl satisfiability and
  model-checking.
\newblock In {\em TACAS 2008, Proceedings}, pages 63--77. Springer, 2008.

\bibitem{Farzan15}
A.~Farzan, Z.~Kincaid, and A.~Podelski.
\newblock Proof spaces for unbounded parallelism.
\newblock {\em SIGPLAN Not.}, 50(1):407--420, Jan. 2015.

\bibitem{Grebenshchikov12}
S.~Grebenshchikov, N.~P. Lopes, C.~Popeea, and A.~Rybalchenko.
\newblock Synthesizing software verifiers from proof rules.
\newblock {\em SIGPLAN Not.}, 47(6):405--416, June 2012.

\bibitem{HJMS02}
T.~A. Henzinger, R.~Jhala, R.~Majumdar, and G.~Sutre.
\newblock Lazy abstraction.
\newblock {\em SIGPLAN Not.}, 37(1):58--70, Jan. 2002.

\bibitem{henzinger:RealTimeSystems}
T.~A. Henzinger, X.~Nicollin, J.~Sifakis, and S.~Yovine.
\newblock Symbolic model checking for real-time systems.
\newblock {\em Information and Computation}, 111:394--406, 1992.

\bibitem{Hoder12}
K.~Hoder and N.~Bj{\o}rner.
\newblock Generalized property directed reachability.
\newblock In {\em SAT 2012. Proceedings}, pages 157--171. Springer, 2012.

\bibitem{Tacas16}
R.~Iosif, A.~Rogalewicz, and T.~Vojnar.
\newblock Abstraction refinement and antichains for trace inclusion of infinite
  state systems.
\newblock In {\em {TACAS} 2016, Proceedings}, pages 71--89, 2016.

\bibitem{Kaminski94}
M.~Kaminski and N.~Francez.
\newblock Finite-memory automata.
\newblock {\em Theor. Comput. Sci.}, 134(2):329--363, Nov. 1994.

\bibitem{Lasota05}
S.~Lasota and I.~Walukiewicz.
\newblock Alternating timed automata.
\newblock In {\em FOSSACS 2005, Proceedings}, pages 250--265. Springer, 2005.

\bibitem{LINCOLN92}
P.~Lincoln, J.~Mitchell, A.~Scedrov, and N.~Shankar.
\newblock Decision problems for propositional linear logic.
\newblock {\em Annals of Pure and Applied Logic}, 56(1):239 -- 311, 1992.

\bibitem{mcmillan06}
K.~L. McMillan.
\newblock Lazy abstraction with interpolants.
\newblock In {\em Proc. of CAV'06}, volume 4144 of {\em LNCS}. Springer, 2006.

\bibitem{McMillan14}
K.~L. McMillan.
\newblock Lazy annotation revisited.
\newblock In {\em CAV2014, Proceedings}, pages 243--259. Springer International
  Publishing, 2014.

\bibitem{ntslib}
{Numerical Transition Systems Repository}.
\newblock \url{http://http://nts.imag.fr/index.php/Flata}, 2012.

\bibitem{OuaknineWorrell04}
J.~Ouaknine and J.~Worrell.
\newblock On the language inclusion problem for timed automata: closing a
  decidability gap.
\newblock In {\em Proceedings of LICS 2004.}, pages 54--63, 2004.

\bibitem{Pnueli77}
A.~Pnueli.
\newblock The temporal logic of programs.
\newblock In {\em Proceedings of the 18th Annual Symposium on Foundations of
  Computer Science}, SFCS '77, pages 46--57. IEEE, 1977.

\bibitem{smrcka}
A.~Smrcka and T.~Vojnar.
\newblock Verifying parametrised hardware designs via counter automata.
\newblock In {\em {HVC}'07}, pages 51--68, 2007.

\bibitem{VARDI94}
M.~Vardi and P.~Wolper.
\newblock Reasoning about infinite computations.
\newblock {\em Information and Computation}, 115(1):1 -- 37, 1994.

\bibitem{symbTransd:POPL12}
M.~Veanes, P.~Hooimeijer, B.~Livshits, D.~Molnar, and N.~Bjorner.
\newblock Symbolic finite state transducers: Algorithms and applications.
\newblock In {\em Proc. of POPL'12}. ACM, 2012.

\bibitem{abp}
A.~Zbrzezny and A.~Polrola.
\newblock Sat-based reachability checking for timed automata with discrete
  data.
\newblock {\em Fundamenta Informaticae}, 79:1--15, 2007.

\end{thebibliography}

\appendix

\subsection{Proof of Lemma \ref{lemma:closure}}

\begin{proposition}\label{prop:overline}
  Given a formula $\phi \in \Form^+(Q,\vec{x})$ and a valuation $\nu$
  mapping each $q \in Q$ to a value $\nu(q) \in \booli$ and each $x \in
  \vec{x}$ to a value $\nu(x) \in \Data^\I$, let $\nu'$ be the
  valuation that assigns each $q \in Q$ the value $\neg\nu(q)$ and
  each $x \in \vec{x}$ the value $\nu(x)$. Then we have 
  $\I,\nu \models \phi$ if and only if $\I,\nu' \not\models \overline{\phi}$. 
\end{proposition}
\proof{Immediate, by induction on the structure of $\phi$. \qed}

\vspace*{\baselineskip}
\proof{ We prove $L(\A_\cup) = L(\A_1) \cup L(\A_2)$ first, the proof
  for $\A_\cap$ being analogous. Let $w = (a_1,\nu_1) \ldots
  (a_n,\nu_n)$ be a word, where $n=0$ corresponds to the empty
  word. We prove by induction on $n\geq0$ that $\Delta(\iota_1 \vee
  \iota_2,a_1\ldots a_n) \iffeq \Delta(\iota_1,a_1\ldots a_n) \vee
  \Delta(\iota_2,a_1\ldots a_n)$. The case $n=0$ follows from the
  definition of the initial configuration of $\A_\cup$. For the
  inductive step $n>0$, $\Delta(\iota_1 \vee \iota_2,a_1\ldots a_n)$
  is obtained from $\Delta(\iota_1 \vee \iota_2,a_1\ldots a_{n-1})$ by
  replacing each variable $q \in \fv{\Bool}{\Delta(\iota_1 \vee
    \iota_2,a_1\ldots a_{n-1})}$ with
  $\Delta(q,a_n)[\vec{x}_{n-1}/\overline{\vec{x}},\vec{x}_n/\vec{x}]$,
  denoted $\Delta^n(\Delta(\iota_1 \vee \iota_2,a_1\ldots
  a_{n-1}),a_n)$. Since by induction hypothesis, $\Delta(\iota_1 \vee
  \iota_2,a_1\ldots a_{n-1}) \iffeq \Delta(\iota_1,a_1\ldots a_{n-1})
  \vee \Delta(\iota_2,a_1\ldots a_{n-1})$, we obtain:
  \[\begin{array}{ll}
  \Delta^n(\Delta(\iota_1 \vee \iota_2,a_1\ldots a_{n-1}),a_n) & \iffeq \\
  \Delta^n(\Delta(\iota_1,a_1\ldots a_{n-1}),a_n) \vee \Delta^n(\Delta(\iota_2,a_1\ldots a_{n-1}),a_n) & \iffeq \\
  \Delta(\iota_1,a_1\ldots a_{n}) \vee \Delta(\iota_2,a_1\ldots a_{n}) \enspace.
  \end{array}\] 
  To prove $L(\overline{\A_1}) = \Sigma[\vec{x}]^* \setminus L(\A_1)$,
  let $w = (a_1,\nu_1) \ldots (a_n,\nu_n)$ be a word and show that
  $\overline{\Delta}(\overline{\iota_1},a_1 \ldots a_n) =
  \overline{\Delta(\iota_1,a_1 \ldots a_n)}$ by induction on $n\geq0$.

  In the case $n=0$, we have $\overline{\Delta}(\overline{\iota_1},a_1
  \ldots a_n) \equiv \overline{\iota_1}$. Then $\varepsilon$ is
  accepted by $\A_1$ iff $\nu_0 \models \iota_1$, where $\nu_0(q) =
  \top$ if $q \in F_1$ and $\nu_0(q) = \bot$, otherwise. But $\nu_0
  \models \iota_1$ iff $\overline{\nu_0} \models \overline{\iota_1}$,
  where $\overline{\nu_0}(q) = \top$ if $q \not\in F_1$ and
  $\overline{\nu_0}(q) = \bot$, otherwise. Thus $\varepsilon$ is
  accepted by $\A_1$ iff it is not accepted by $\overline{\A_1}$.

  For the case $n>0$, we compute:
  \[\begin{array}{lcl}
  \overline{\Delta}^n(\overline{\Delta}(\iota_1,a_1\ldots a_{n-1}),a_n) & \iffeq & \text{ (by ind. hyp.)} \\
  \overline{\Delta}^n(\overline{\Delta(\iota_1,a_1\ldots a_{n-1})},a_n) & \iffeq & \text{ (by the def. of $\overline{\phi}$)} \\
  \overline{\Delta(\iota_1,a_1\ldots,a_n)} \enspace.
  \end{array}\]
  Let $\nu,\nu' : (Q \cup \bigcup_{i=0}^n\vec{x}_i) \rightarrow (\booli
  \cup \Data^\I)$ be valuations such that: \begin{compactitem}
  \item $\nu(q) = \top$ and $\nu'(q) = \bot$, for each $q \in F$, 
  \item $\nu(q) = \bot$ and $\nu'(q) = \top$, for each $q \in Q \setminus F$, 
  \item $\nu(x) = \nu'(x)$, for each $x \in \vec{x}_0$, 
  \item $\nu(x) = \nu'(x) = \nu_i(x)$, for each $x \in \vec{x}_i$ and each $i
    \in [1,n]$.
  \end{compactitem}
  By Proposition \ref{prop:overline}, we have $\I,\nu \models
  \Delta(\iota_1,a_1 \ldots a_n) \iff \I,\nu' \not\models
  \overline{\Delta(\iota_1,a_1 \ldots a_n)} \iff \I,\nu' \not\models
  \overline{\Delta}(\iota_1,a_1 \ldots a_n)$. Thus for all $w \in
  \Sigma[\vec{x}]^*$, we have $w \in L(\A_1) \iff w \not\in
  L(\overline{\A_1})$. \qed}

\subsection{Proof of Lemma \ref{lemma:safety-invariant}}

\proof{ Let $\A = \tuple{\vec{x},Q,\iota,F,\Delta}$ in the following.
  ``$\Leftarrow$'' This direction is trivial. ``$\Rightarrow$'' We
  define $\Inv : (Q \mapsto \booli) \rightarrow 2^{\vec{x} \mapsto
    \Data^\I}$ as follows. For each $\beta : Q \rightarrow \booli$,
  let $\Inv(\beta) = \{\nu : \vec{x} \rightarrow \Data^\I \mid \exists
  u \in \Sigma^* ~.~ \beta\cup\nu \models
  \Post{\A}(\iota,u)\}$. Checking that $\Inv$ is a safety invariant is
  straightforward. \qed}

\subsection{Proof of Lemma \ref{lemma:abstract-invariant}}

\proof{ It is sufficient to show that there exists $k\geq0$ such that for
  all $u \in \Sigma^*$ there exists $i \in [0,k]$ such that
  $\Post{\A}(\iota,u) \models \AbsPost{\A}(\iota,\mu(i))$. We have
  $\Post{\A}(\iota,u) \models \AbsPost{\A}(\iota,u)$ for all $u \in
  \Sigma^*$. But since $\preds$ is a finite set, also the set
  $\{\AbsPost{\A}(\iota,u) \mid u \in \Sigma^*\}$ is finite. Thus
  there exists $k \geq 0$ such that, for all $u \in \Sigma^*$ there
  exists $i \in [0,k]$ such that $\AbsPost{\A}(\iota,u) \iffeq
  \AbsPost{\A}(\iota,\mu(i))$, which concludes the proof. \qed}

\subsection{Proof of Proposition \ref{prop:positive-interpol}}

\begin{proposition}\label{prop:asterisk}
  Given a formula $\phi \in \Form^+(Q,\vec{x})$ and $a \in \Sigma$, we
  have \(\Delta(\phi,a) \iffeq \exists Q' ~.~ \phi[Q'/Q] \wedge
  \bigwedge_{q \in Q} (q' \rightarrow \Delta(q,a))\).
\end{proposition}
\proof{``$\Rightarrow$'' If $\I,\beta\cup\overline\nu\cup\nu \models
  \Delta(\phi,a)$, for some valuations $\beta : Q \rightarrow \booli$
  and $\overline\nu : \overline{\vec{x}} \rightarrow \Data^\I$, $\nu :
  \vec{x} \rightarrow \Data^\I$, then we build a valuation $\beta' :
  Q' \rightarrow \booli$ such that
  $\I,\beta'\cup\beta\cup\overline\nu\cup\nu \models \phi[Q'/Q] \wedge
  \bigwedge_{q \in Q} (q' \rightarrow \Delta(q,a))$. For each
  occurrence of a formula $\Delta(q,a)$ in $\Delta(\phi,a)$ we set
  $\beta'(q') = \true$ if $\I,\beta\cup\overline\nu\cup\nu \models
  \Delta(q,a)$ and $\beta'(q') = \false$, otherwise. Since there are
  no negated occurrences of such subformulae, the definition of
  $\beta'$ is consistent, and the check
  $\I,\beta'\cup\beta\cup\overline\nu\cup\nu \models \phi[Q'/Q] \wedge
  \bigwedge_{q \in Q} (q' \rightarrow \Delta(q,a))$ is
  immediate. $''\Leftarrow''$ This direction is an easy check. \qed}

\vspace*{\baselineskip} \proof{ Let
  $\tuple{\top,I_0,\ldots,I_{n},\bot}$ be an interpolant, and $i \in
  [0,n]$ be the first index for which $I_i \not\in
  \Form^+(Q,\vec{x})$. If $i=0$, we replace $I_0$ with $\iota \in
  \Form^+(Q,\vec{x})$ and verify that the result is still an
  interpolant. If $i>0$, by Proposition \ref{prop:asterisk}, we have
  that $\Delta^{i-1}(I_{i-1},a_i) \iffeq \exists Q_{i-1} ~.~
  I_{i-1}[Q_{i-1}/Q] \wedge \theta_i$. Since $I_{i-1} \in
  \Form^+(Q,\vec{x})$, by the choice of $i$, we have that
  $\Delta^{i-1}(I_{i-1},a_i) \in \Form^+(Q,\vec{x})$ as well. We
  replace $I_i$ with $\Delta^{i-1}(I_{i-1},a_i)$. The result is still
  an interpolant, because: \begin{compactitem}
  \item $I_{i-1}[Q_{i-1}/Q] \wedge \theta_i \models
    \Delta^{i-1}(I_{i-1},a_i)$, by Proposition \ref{prop:asterisk}, and
  \item $\Delta^{i-1}(I_{i-1},a_i) \wedge \theta_{i+1} \models \exists
    Q_{i-1} ~.~ I_{i-1}[Q_{i-1}/Q] \wedge \theta_i \wedge \theta_{i+1}
    \models I_i[Q_i/Q] \wedge \theta_{i+1} \models I_{i+1}[Q_{i+1}/Q]$.
  \end{compactitem}
  We proceed with these replacements until there are no more formulae
  $I_i \not\in \Form^+(Q,\vec{x})$ left. \qed}

\subsection{Proof of Lemma \ref{lemma:progress}}

\proof{ Let $\Theta(u) \equiv \theta_0(Q_0) \wedge \theta_1(Q_0 \cup
  Q_1,\vec{x}_0 \cup \vec{x}_1) \wedge \ldots \wedge \theta_n(Q_{n-1}
  \cup Q_n,\vec{x}_{n-1} \cup \vec{x}_n) \wedge \theta_{n+1}(Q_n)$ in
  the following.

  (\ref{it1:progress}) We apply Proposition \ref{prop:asterisk}
  recursively and get: \[\Post{\A}(\iota,u)[Q_n/Q,\vec{x}_n/\vec{x}]
  \iff \exists Q_0 \ldots \exists Q_{n-1} \exists \vec{x}_0 \ldots
  \exists \vec{x}_{n-1} ~.~ \bigwedge_{i=0}^n \theta_i\] Assuming that
  $\Theta(u)$ is satisfiable, we obtain a model for $\Accept{\A}(u)
  \equiv \Post{\A}(\iota,u) \wedge \theta_{n+1}[Q/Q_n]$.

  (\ref{it2:progress}) If $\tuple{\top,I_0,\dots,I_n,\bot}$ is an
  interpolant for $\Theta(u)$, the following entailments
  hold: \begin{compactitem}
  \item $\theta_0 \models I_0[Q_0/Q]$,
  \item $I_{k-1}[Q_{k-1}/Q,\vec{x}_{k-1}/\vec{x}] \wedge \theta_k \models I_{k}[Q_k/Q,\vec{x}_k/\vec{x}]$, $\forall k\in[1,n]$. 
  \item $I_{n}[Q_{n}/Q] \wedge \theta_{n+1} \models \bot$. 
  \end{compactitem}
  We prove that $\AbsPost{\A}(\iota,a_1\ldots a_n) \models I_{n}$ by
  induction on $n \geq 0$. This is sufficient to conclude because
  $\AbsAccept{\A}(a_1\ldots a_n) \equiv \AbsPost{\A}(\iota,a_1\ldots
  a_n) \wedge \theta_{n+1}[Q/Q_n] \models I_{n} \wedge
  \theta_{n+1}[Q/Q_n] \models \bot$. For the base case $n=0$, we have
  $\AbsPost{\A}(\iota,\varepsilon) \equiv \iota \equiv \theta_0[Q/Q_0]
  \models I_0$. For the induction step $n>0$, we compute:
  \[\begin{array}{l}
  \AbsPost{\A}(\iota,a_1\ldots a_n)[Q_n/Q] \equiv \text{ (by def. of $\AbsPost{\A}$)} \\
  \exists \vec{x}_{n-1} ~.~ \abs{\Delta^n(\AbsPost{\A}(\iota,a_1\ldots a_{n-1}),a_n)}[Q_n/Q] \models \text{ (by Prop. \ref{prop:asterisk})} \\
  \exists Q_{n-1}  \exists \vec{x}_{n-1} ~.~ \AbsPost{\A}(\iota,a_1\ldots a_{n-1})[Q_{n-1}/Q] \wedge \theta_n \models \text{ (ind. hyp.) } \\
  \exists Q_{n-1}  \exists \vec{x}_{n-1} ~.~ I_{n-1}[Q_{n-1}/Q] \wedge \theta_n \models I_n[Q_n/Q] \hfill\text{\qed}
  \end{array}\]}

\subsection{Proof of Theorem \ref{thm:predabs}}

\proof{We prove the following invariant: each time Algorithm
  \ref{alg:predabs} reaches line \ref{ln:while}, the set $W$ of nodes
  in $\worklist$ contains all the frontier nodes in the ART
  $\tuple{N\cup W,E,\rootNode,\Lambda,\lhd}$ which are not covered by
  some node in $N$, namely that:
  \begin{equation}\label{inv:predabs}
  W = \set{n \mid \forall m \in N~ \forall a \in \Sigma ~.~ (n,a,m)
    \not\in E ~\wedge~ (n,m) \not\in \lhd}
  \end{equation}
  Initially, this is the case because $W = \set{\rootNode}$ and $E =
  \lhd = \emptyset$. If the invariant holds previously, at line
  \ref{ln:while}, it will hold again after line \ref{ln:refine} is
  executed, because, when the subtree rooted at the pivot $p$ is
  removed, $p$ becomes a member of the set of uncovered frontier
  nodes, and is added to $W$ at line \ref{ln:refine}. Otherwise, the
  invariant holds at line \ref{ln:while} and the control follows the
  else branch at line \ref{ln:else}. In this case, the newly created
  frontier node $s$ is added to $W$ only if it is not covered by an
  existing node in $N$ (line \ref{ln:covered}).

  Next we prove that, if Algorithm \ref{alg:predabs} returns $\true$,
  then $\bigvee_{n \in N} \Lambda(n)$ defines a safety
  invariant. Suppose that Algorithm \ref{alg:predabs} returns at line
  \ref{ln:true}. Then it must be that $W = \emptyset$. Because
  (\ref{inv:predabs}) is invariant, each node in $N$ is either covered
  by another node in $N$, or all its successors are in $N$. We prove
  first that $\bigvee_{n \in N} \Lambda(n)$ is an invariant: for any
  $u \in \Sigma^*$, there exists some node $n \in N$ such that
  $\Post{\A}(\iota,u) \models \Lambda(n)$. Let $u \in \Sigma^*$ be an
  arbitrary sequence. If $u$ labels the path from $\rootNode$ to some
  $n \in N$, we have $\Post{\A}(\iota,u) \models \AbsPost{\A}(\iota,u)
  \models \Lambda(n)$ and we are done. Otherwise, let $v$ be the
  (possibly empty) prefix of $u$ which labels the path from
  $\rootNode$ to some $n \in N$, which is covered by another $m \in
  N$, where $(n,a,m) \in E$, that is $u = vav'$, for some $a \in
  \Sigma$ and $v'\in\Sigma^*$. Moreover, we have $\Post{\A}(\iota,va)
  \models \AbsPost{\A}(\iota,va) \models \Lambda(m)$, by the
  construction of the set $\lhd$ of covering edges --- lines
  \ref{ln:direct-cover}, \ref{ln:child-cover},
  \ref{ln:indirect-cover}. Continuing this argument recursively from
  $m$, since $\len{v'} < \len{u}$, we shall eventually discover a node
  $p$ such that $\Post{\A}(\iota,u) \models \Lambda(p)$.

  To prove that $\bigvee_{n \in N} \Lambda(n)$ is, moreover, a safety
  invariant, suppose, by contradiction, that there exists $u \in
  \Sigma^*$ such that $\Accept{\A}(u)$ is satisfiable. By the previous
  point, there exists a node $p \in N$ such that $\Post{\A}(\iota,u)
  \models \Lambda(p)$. But then we have $\Accept{\A}(\iota,u) \models
  \Accept{\A}(\iota,\lambda(p))$, thus $\Accept{\A}(\iota,\lambda(p))$
  is satisfiable as well. However, this cannot be the case, because
  $p$ has been processed at line \ref{ln:feasible} and Algorithm
  \ref{alg:predabs} would have returned a counterexample,
  contradicting the assumption that it returns $\true$. This concludes
  the proof that $\bigvee_{n\in N} \Lambda(n)$ is a safety invariant,
  thus $L(\A) = \emptyset$, by Lemma \ref{lemma:safety-invariant}. We
  have then proved the second point of the statement.

  For the first point, assume that $L(\A) \neq 0$ and let $w =
  (a_1,\nu_1) \ldots (a_k,\nu_k) \in L(\A)$ be a word. By the above,
  Algorithm \ref{alg:predabs} cannot return $\true$. Suppose, by
  contradiction that it does not terminate. Since the sequences from
  $\Sigma^*$ are explored in breadth-first order, every sequence of
  length $k$ is eventually explored, which leads to the discovery of
  $w$ at line \ref{ln:feasible}. Then Algorithm \ref{alg:predabs}
  terminates returning $w \in L(\A)$. \qed}

\subsection{Proof of Theorem \ref{thm:impact}}

\begin{lemma}\label{lemma:well-labeled}
  Given an ART $\Art = \tuple{N,E,\rootNode,\Lambda,R,T,\lhd}$ built
  by Algorithm \ref{alg:impact}, $\Post{\A}(\Lambda(n),a) \models
  \Lambda(m)$, for all $(n,a,m) \in E$.
\end{lemma}
\proof{ We distinguish two cases. First, if $(n,a,m)$ occurs on a path
  in $\Art$ that has never been refined, then $\Lambda(m)=\top$ and
  the entailment holds trivially. Otherwise, let $\Omega$ be the set
  of paths $\omega = (n_0,a_1,n_1), \ldots, (n_{k-1},a_k,n_k)$, where
  $n_0 = \rootNode$ and $(n,a,m) = (n_{i-1},a_i,n_i)$, for some $i \in
  [1,k]$ and, moreover, $a_1\ldots a_k$ was found, at some point, to
  be a spurious counterexample. Let
  $\langle\top,I^\omega_0,\ldots,I^\omega_k,\bot\rangle$ be an
  interpolant for $\Phi(a_1 \ldots a_k) \equiv \Lambda(\rootNode)
  \wedge \bigwedge_{i=1}^k \theta_i \wedge \bigwedge_{q \in R(n_k)}
  (q_k \rightarrow \bot)$, such that $I^\omega_i \in
  \Form^+(Q,\vec{x})$, for all $i \in [0,k]$. According to Proposition
  \ref{prop:positive-interpol}, it is possible to build such an
  interpolant, when $\Phi(a_1 \ldots a_k)$ is unsatisfiable. By
  Proposition \ref{prop:asterisk}, we obtain
  $\Delta^i(I^\omega_{i-1},a_i)[Q_i/Q] \iffeq \exists Q_{i-1} ~.~
  I^\omega_{i-1}[Q_{i-1}/Q,\vec{x}_{i-1}/\vec{x}] \wedge \theta_i$
  and, since $I^\omega_{i-1}[Q_{i-1}/Q,\vec{x}_{i-1}/\vec{x}] \wedge
  \theta_i \models I^\omega_i[Q_i/Q,\vec{x}_i/\vec{x}]$, we obtain
  that $\Delta^i(I^\omega_{i-1},a_i)[Q_i/Q] \models
  I^\omega_i[Q_i/Q,\vec{x}_i/\vec{x}]$. Since $\Lambda(n_{i-1}) =
  \bigwedge_{\omega \in \Omega} I^\omega_{i-1}$ and $\Lambda(n_{i}) =
  \bigwedge_{\omega \in \Omega} I^\omega_{i}$, we obtain
  $\Post{\A}(\Lambda(n_{i-1}),a_i) \models \Lambda(n_i)$. \qed}

\vspace*{\baselineskip}
\proof{ We prove first that, each time Algorithm \ref{alg:impact}
  reaches the line \ref{ln:impact-while}, we have:
  \begin{equation}\label{inv:impact}
  W = \set{n \mid n \text{ uncovered, } \exists a \in \Sigma~ \forall
    s \in N ~.~ (n,a,s) \not\in E}
  \end{equation}
  Initially, $W = \set{\rootNode}$ and $E=\lhd=\emptyset$, thus
  (\ref{inv:impact}) holds trivially. Suppose that (\ref{inv:impact})
  holds at when reaching line \ref{ln:impact-while} and some node $n$
  was removed from $W$ and inserted into $N$. We distinguish two
  cases, either: \begin{itemize}
  \item $n$ is covered, in which case $W$ becomes $W \setminus
    \set{n}$ and (\ref{inv:impact}) holds, or
  \item $n$ is not covered, in which case $W$ becomes $(W \setminus
    \set{n}) \cup S$, where $S = \set{s \not\in N \mid (n,a,s) \in E,
      a \in \Sigma}$ is the set of fresh successors of $n$. But then
    no node $s \in S$ is covered and has successors in $E$, thus
    (\ref{inv:impact}) holds.
  \end{itemize}
  Then the condition (\ref{inv:impact}) holds next time line
  \ref{ln:impact-while} is reached, thus it is invariant. 

  Suppose first that Algorithm \ref{alg:impact} returns $\true$, thus
  $W = \emptyset$ and, by (\ref{inv:impact}), for each node in $n \in
  N$ one of the following hold: \begin{compactitem}
  \item $n$ is covered, or
  \item for each $a \in \Sigma$ there exists $s \in N$ such that
    $(n,a,s) \in E$. 
  \end{compactitem}
  We prove that, in this case, $\bigvee_{n \in N} \Lambda(n)$ defines
  a safety invariant and conclude that $L(\A) = \emptyset$, by Lemma
  \ref{lemma:safety-invariant}. To this end, let $u = a_1 \ldots a_k
  \in \Sigma^*$ be an arbitrary sequence and let $v_1$ be the largest
  prefix of $u$ that labels a path from $\rootNode$ to some node $n_1
  \in N$. If $v_1 = u$ we are done. Otherwise, by the choice of $v_1$,
  it must be the case that a successor of $n_1$ is missing from
  $(N,E)$, thus $n_1$ must be covered, by (\ref{inv:impact}) and the
  fact that $W=\emptyset$. Let $n'_1$ be the closest ancestor of $n_1$
  such that $(n'_1,n''_1) \in \lhd$, for some $n''_1 \in N$, and let
  $v'_1$ be the prefix of $v_1$ leading to $n'_1$. By the construction
  of $\lhd$ (line \ref{ln:close-uncover} in function
  $\Call{Close}{}$), we have $\Lambda(n'_1) \models
  \Lambda(n''_1)$. Applying Lemma \ref{lemma:well-labeled} inductively
  on $v'_1$, we obtain that $\Post{\A}(\iota,v'_1) \models
  \Lambda(n'_1)$, thus $\Post{\A}(\iota,v'_1) \models \Lambda(n''_1)$.
  Continuing inductively from $n''_1$, we exhibit a sequence of
  strings $v'_1,\ldots,v'_\ell \in \Sigma^*$ and nodes $\rootNode =
  m_0,m_1,\ldots,m_\ell$ such that, for all $i \in
  [1,\ell]$: \begin{compactitem}
  \item $v'_i$ labels the path between $m_{i-1}$ and $m_i$ in $(N,E)$,    
  \item $\Post{\A}(\iota,v'_1\ldots v'_i) \models \Lambda(m_i)$.
  \end{compactitem}
  Moreover, we have $u = v'_1 \ldots v'_\ell$, thus
  $\Post{\A}(\iota,u) \models \Lambda(m_\ell)$ and we are done showing
  that $\bigvee_{n\in N} \Lambda(n)$ is an invariant. 

  To prove that $\bigvee_{n \in N} \Lambda(n)$ is, moreover, a safety
  invariant, suppose that $\Accept{\A}(u)$ is satisfiable, for some $u
  \in \Sigma^*$ and let $n \in N$ be a node such that
  $\Post{\A}(\iota,u) \models \Lambda(n)$. By the previous point, such
  a node must exist. But then $\Accept{\A}(u) \models
  \Accept{\A}(\lambda(n))$, thus $\Accept{\A}(\lambda(n))$ is
  satisfiable, and Algorithm \ref{alg:impact} returns at line
  \ref{ln:impact-nonempty}, upon encountering $\lambda(n)$. But this
  contradicts the assumption that Algorithm \ref{alg:impact} returns
  $\true$, hence we have proved that $\bigvee_{n \in N} \Lambda(n)$ is
  a safety invariant, and $L(\A) = \emptyset$ follows, by Lemma
  \ref{lemma:safety-invariant}. We have then proved the second point
  of the statement.

  To prove the first point, assume that $L(\A) \neq \emptyset$. By the
  previous point, Algorithm \ref{alg:impact} does not return
  $\true$. Suppose, by contradiction, that it does not terminate and
  conclude using the breadth-first argument from the proof of Theorem
  \ref{thm:predabs}. \qed}

\end{document}